\documentclass[11pt,preprint,english,fleqn,authoryear]{elsarticle}

\usepackage{ae,aecompl}
\usepackage[T1]{fontenc}
\usepackage[latin9]{inputenc}
\usepackage{geometry}
\makeatletter
\providecommand{\tabularnewline}{\\}

\usepackage{float}
\usepackage{amsthm}
\usepackage{amsmath}
\usepackage{tikz}
\usepackage{algorithmic}
\usepackage{breqn}
\usepackage{bm}
\usepackage{verbatim}
\usepackage{xcolor}
\usepackage{graphicx}
\usepackage{caption}
\usepackage[labelformat=simple]{subcaption}

\usepackage{babel}
\usepackage{comment}

\usepackage{enumitem}

\usepackage{lineno}
\usepackage{multirow}

\usepackage{framed}
\usepackage{longtable}
\usepackage{siunitx}

\addto\captionsenglish{}

\definecolor{wine-stain}{rgb}{0.5,0,0}
\usepackage{hyperref}
\hypersetup{colorlinks=true,citecolor=blue,linktoc=all, linktocpage=false, linkcolor=wine-stain}

\def\ps@pprintTitle{%
	\let\@oddhead\@empty
	\let\@evenhead\@empty
	\def\@oddfoot{\centerline{\thepage}}%
	\let\@evenfoot\@oddfoot}
\makeatother

\begin{document}
	
\begin{frontmatter} 
\title{Adiabatic Lapse Rate and Static Stability in the Venus Atmosphere
	calculated from Real Gas Mixture Models}

\author[ssec]{Arkopal Dutt\corref{cor1}\fnref{padd}}
\ead{arkopal@mit.edu}

\author[ssec]{Sanjay S. Limaye}
\ead{sanjayl@ssec.wisc.edu}

\address[ssec]{Space Science and Engineering Center, University of Wisconsin-Madison, Madison, WI 53706, USA}

\cortext[cor1]{Corresponding author}
\fntext[padd]{Present address: Department of Mechanical Engineering, Massachusetts Institute of Technology, Cambridge, MA 02139, USA}
	
\begin{abstract} 
It is known that the ideal gas equation of state is not valid in the lower atmosphere of Venus where surface pressures reach $9$ MPa and surface temperatures approach $750$ K. Moreover, the presence of a small amount of nitrogen slightly complicates the calculation of thermodynamic properties of the real gas mixture present in the atmosphere. Previous calculations of the adiabatic lapse rate in the Venus atmosphere have used approximations to estimate the adiabatic lapse rate. Here, we calculate the adiabatic lapse rate more accurately by using multi-parameter mixture models formulated in reduced Helmholtz free energy to account for the real gas mixture effects. Our results show small differences from the \cite{Seiff1980} values for the adiabatic lapse rate which may be significant where the Venus atmosphere is close to being neutral. For accurate knowledge of the static stability for atmosphere circulation, a local value of the adiabatic lapse rate is necessary.
\end{abstract}

\end{frontmatter}

\section{Introduction} \label{sec:Intro}
\citet{Staley1970} pointed out that the adiabatic lapse rate for the lower atmosphere of Venus cannot be calculated using the ideal gas equation ($g/c_p$) due to the high temperature and pressure conditions and the presence of small amount of nitrogen. Considering an arbitrary equation of state for any gas mixture, \citet{Staley1970} derived the following expression for adiabatic lapse rate $\Gamma$ at the altitude $z$ in a planetary atmosphere
\begin{equation}
\Gamma=-\frac{dT}{dz}=-\frac{T}{\rho}\frac{\left(\frac{\partial p}{\partial T}\right)_{\rho}}{\left(\frac{\partial p}{\partial\rho}\right)_{T}}\left(\frac{g}{c_{p}}\right) = \frac{T}{\rho}\left(\frac{\partial\rho}{\partial T}\right)_{p}\left(\frac{g}{c_{p}}\right) 
\label{eq:adiabatic_lapse_rate}
\end{equation}
where $T$ is temperature, $p$ is pressure, $\rho$ is the density, $g$ is the acceleration due to gravity and $c_p$ is the isobaric specific heat capacity of the air at altitude $z$. Assuming that the atmosphere is composed of pure $CO_{2}$, \citet{Staley1970} calculated $c_{p}$ and hence $\Gamma$ using the real gas physical properties of pure $CO_{2}$ \citep{Hilsenrath1955} across a range of pressure and temperature that can be found in the atmosphere of Venus. The major shortcoming of this approach was that the presence of $N_{2}$ in the atmosphere was neglected.
In order to overcome this, \citet{Seiff1980} calculated the adiabatic lapse rate by assuming an ideal binary gas mixture of real gas components: carbon dioxide ($CO_{2}$) and nitrogen ($N_{2}$) in a volume measured mixing ratio of $96.5:3.5$, arguing that the abundance of nitrogen is small. In this approach, adiabatic lapse rate is written as
\begin{equation}
\Gamma=-(aT)\frac{g}{c_{p}}
\label{eq:adiabatic_lapse_rate_Seiff}
\end{equation}
where 
\begin{equation}
a=-\frac{1}{\rho}\left(\frac{\partial\rho}{\partial T}\right)_{p}
\label{eq:a_Seiff}
\end{equation}
In the case of an ideal binary gas mixture, the contribution of pure real gas component $i$ to the thermodynamic properties of the mixture is directly proportional to it's mole fraction $x_{i}$ which gives
\begin{align}
c_{p} & = \sum_{i}x_{i}c_{pi}\label{cp_Seiff}\\
aT & = \sum_{i}x_{i}(aT)_{i}
\label{eq:combination_rules_Seiff}
\end{align}
The main drawback of this method is that the non-ideal interactions of $CO_{2}$ and $N_{2}$ in the mixture are neglected in calculating the thermodynamic properties of the mixture.

Furthermore, the VIRA model \citep{Seiff1985} extrapolated the surface temperature of Venus below 12 km altitude (at which the last measurements were made by the sensors on the four Pioneer probes) by using the adiabatic lapse rate calculated by \citet{Seiff1980}. Thus surface temperatures reported for all Pioneer probes are slightly inaccurate. As a result, the values calculated for the surface conditions on Venus which has been used in most subsequent studies pertaining to the stability of atmosphere and atmospheric circulation can be made more accurate. VeGa2 lander is the only atmospheric probe which has provided us with accurate measurements down to the surface. VeGa2 lander data in the lower atmosphere was examined by \citet{Seiff1987} which showed near neutral and superadiabatic layers. Presence of superadiabatic layers on Venus raises some key questions about the source of near surface heat deposition and the resulting atmospheric circulation in the lower atmosphere. An important parameter in understanding such atmospheric processes on Venus is static stability which influences small-scale turbulence caused by convection or wind shear, mesoscale motions and large-scale circulations as well as topography induced disturbances by the ambient flow. Thus is it imperative to calculate the adiabatic lapse rate accurately for the known conditions on Venus. 

A more detailed derivation of the real gas adiabatic lapse rate for a planetary atmosphere with a multi-component real gas mixture composition varying with altitude is illustrated in \ref{Deriv_Gamma}. The same expressions for adiabatic lapse rate (Eqs.\ref{eq:adiabatic_lapse_rate}) as originally derived by \citet{Staley1970} are obtained. As can be seen from the expressions, accuracy in adiabatic lapse rate at any altitude depends on the accuracy in Venus atmosphere profiles available, composition of the atmosphere at that altitude, density $\rho$ and the isobaric specific heat capacity $c_{p}$ of air at that altitude. As there is limited experimental data available for the particular real gas binary mixture that largely makes up the Venus atmosphere, it becomes necessary to use an equation of state to predict the density $\rho$ and isobaric specific heat capacity $c_{p}$ at different pressures $p$ and temperatures $T$. We have already highlighted how the approaches followed by both \citet{Staley1970} and \citet{Seiff1980} introduced errors in the determination of these quantities for the real gas binary mixture of $CO_{2}-N_{2}$ that largely make up the Venus atmosphere.

In this work, we determine $\rho$ and $c_{p}$ more accurately than previous approaches by considering the interactions between real gas components in the mixture through a equation of state for the mixture. A variety of different equations of state for fluids and mixtures exist \citep{Sengers2000}. Here, we determine the physical properties of the real gas binary mixture $CO_{2}-N_{2}$ by using thermodynamic models in Helmholtz energy. We consider two different Helmholtz energy mixture models proposed in \citep{Lemmon1999} and \citep{Kunz2012}. The advantage these models present over other equations of state for mixtures is that it allows us to obtain the mixture properties by combining properties of real gas components obtained through their respective equations of state. In Sections \ref{sec:HEOS_Models} and \ref{sec:Density_Solvers}, we review these models and how they can be used to calculate the desired thermodynamic quantities. This will be followed by a verification of the approach against experimental data and prior approaches in Section \ref{sec:Verification-against-Experiments}. In Section \ref{sec:Adiabatic-Lapse-Rate}, we show how the mixture models can be used to calculate the adiabatic lapse rate and static stability for the Venus atmosphere. Results are discussed in Section \ref{sec:Results}. Finally in Section \ref{sec:Conclusion-and-Future}, we highlight the results obtained and discuss future work.

\section{Background} \label{sec:HEOS_Models}
\subsection{Review of Mixture Models}
Equations of state formulated in reduced Helmholtz free energy for mixtures were first proposed independently by \cite{tillner1993} and \cite{Lemmon1996}. These empirical multi-parameter models rely on mixing rules to obtain properties of multi-component mixtures from equations of states of the pure fluid components. These mixing rules and the equations of state of the pure fluid components themselves are obtained through fitting of experimental data of multiple thermodynamic properties. The first mixture model that we consider was proposed in \citep{Lemmon1999}. The second mixture model that we consider is the GERG-2008 model which was proposed in \citep{Kunz2012} and is considered the most accurate mixture model for obtaining thermodynamic properties of natural gases. The older GERG-2004 mixture model \citep{Kunz2004} was used by \cite{hagermann2007speed} to estimate the abundance of methane in Titan's atmosphere using speed of sound measurements through a Bayesian analysis. However, mixture models in Helmholtz energy have not been applied to compute adiabatic lapse rate of a multi-component planetary atmosphere before. 

A more recent mixture model was proposed in \citep{gernert2013new} to predict thermodynamic properties mixtures relevant for Carbon Capture Storage more accurately. However, the mixing rule suggested for the binary mixture of carbon dioxide and nitrogen in \citep{Kunz2012} remains unchanged. We consider the two different mixture models to reflect the effect of mixing rules on accuracy even when both models use the same pure fluid equations of state.  From here on, we will refer to the mixture model introduced in \citep{Lemmon1999} as LJ-1999 model and \cite{Seiff1980} approach of considering ideal mixture of real gases as IMRG model.
\subsection{Mixture Model in Helmholtz Free Energy}
Any generalized mixture model in Helmholtz free energy $A$ with independent mixture variables $\rho$, temperature $T$ and molar composition $\bar{x}$ \citep{Lemmon1996, Lemmon1999, Kunz2012} can be written as
\begin{equation}
A(\tilde{\rho},T,\bar{x}) = A^{idmix}(\tilde{\rho},T,\bar{x}) + A^{E}(\tilde{rho},T,\bar{x})\label{eq:mix_lemmon}
\end{equation}
where $A^{idmix}$ is the Helmholtz energy of the ideal mixture of the real gas components, $A^{E}$ is the Helmholtz energy contribution to mixing, $\tilde{\rho}$ ($=\rho/M$) is the amount of substance density and $M$ is the molar mass of the mixture. In general, $M=\sum_{i}x_{i}M_{i}$ where $M_{i}$ is the molar mass of component $i$. \citet{Seiff1980} essentially neglected $A^{E}$ in calculating $c_p$ in Eq.\ref{cp_Seiff}. It is however easier to work with the following decomposition of the Helmholtz energy of the mixture
\begin{equation}
A(\tilde{\rho},T,\bar{x}) = A^{o}(\tilde{\rho},T,\bar{x}) + A^{r}(\tilde{\rho},T,\bar{x})\label{eq:mix_lemmon2}
\end{equation}
where $A^{o}$ is the contribution of the ideal gas and $A^{r}$ is the contribution from the residual Helmholtz energy of the pure fluid components and from the Helmholtz energy contribution to mixing. Non-dimensionalizing Eq.\ref{eq:mix_lemmon2} by dividing by $RT$ ($R=8.314510\,\text{J/(mol\ensuremath{\cdot}K)}$ is the universal gas constant and $T$ is the mixture temperature), we obtain
\begin{equation}
\alpha(\tilde{\rho},T,\bar{x}) = \alpha^{o}(\tilde{\rho},T,\bar{x}) + \alpha^{r}(\delta,\tau,\bar{x})
\end{equation}
where $\delta$ is the reduced mixture density and $\tau$ is the inverse reduced mixture temperature given by
\begin{eqnarray}
\delta & = & \tilde{\rho}/\rho_r(\bar{x})	\\
\tau & = & T_r(\bar{x})/T
\end{eqnarray}
These reducing parameters are only functions of the composition as indicated above. They are specific to the mixing rule that is followed. For example, the reducing function used in the Lemmon's model \citep{Lemmon1999} is very different from that used in the GERG-2008 model \citep{Kunz2012}. The non-dimensionalized Helmholtz free energy of the ideal gas mixture is
\begin{equation}
\frac{A^o(\tilde{\rho},T,\bar{x})}{RT} = \alpha^{o}(\tilde{\rho},T,\bar{x}) = \sum \limits_{i=1}^{n}x_{i}\left[\alpha_{i}^{o}(\tilde{\rho},T)+\ln x_{i}\right]\label{eq:alpha_o_lemmon}
\end{equation}
where $\alpha_{i}^{o}$ is the ideal gas Helmholtz energy of component $i$ in the mixture which is a function of the mixture amount of substance density $\tilde{\rho}$ and temperature $T$, and not that of reduced density $\delta$ and inverse reduced temperature $\tau$. The term $\sum \limits_{i=1}^{n}x_{i}\ln x_{i}$ quantifies the entropy of mixing. The residual part of the non-dimensionalized Helmholtz free energy is
\begin{equation}
\frac{A^r}{RT} = \alpha^{r} = \sum \limits_{i=1}^{n}x_{i}\alpha_{i}^{r}(\delta,\tau) + \alpha^{E}(\delta,\tau,\bar{x})\label{eq:alpha_r_lemmon}
\end{equation}
where $\alpha_i^r$ is the non-dimensionalized residual part of Helmholtz free energy of component $i$ in the mixture and $\alpha_E$ is called the excess value of the non-dimensionalized Helmholtz free energy or the departure function \citep{Kunz2012}. The usual functional form is 
\begin{equation}
	\alpha^E(\delta,\tau,\bar{x}) = \sum \limits_{i=1}^{n-1} \sum \limits_{j=i+1}^{n} x_i x_j F_{ij} \alpha^{r}_{ij}(\delta,\tau)
	\label{eq:func_alphaE_HEOS}
\end{equation}
where the functional form of $\alpha^{r}_{ij}$ and the value of parameter $F_{ij}$ is prescribed by the mixing rule being used.
All common thermodynamic properties such as pressure, isochoric heat capacity, isobaric heat capacity, sound of speed, enthalpy, saturated-liquid density and VLE data can be obtained from the derivatives of $\alpha^{0}$ and $\alpha^{r}$. A list of the expressions can be found in \citep{Kunz2012}. Here, we only list those that are of relevance to us
\begin{equation}
p=\tilde{\rho}RT\left[1+\delta\left(\frac{\partial\alpha^{r}}{\partial\delta}\right)_{\tau}\right]
\label{eq:p_HEOS}
\end{equation}
\begin{equation}
\frac{\tilde{c}_{v}}{R}=-\tau^{2}\left[\left(\frac{\partial^{2}\alpha^{0}}{\partial\tau^{2}}\right)+\left(\frac{\partial^{2}\alpha^{r}}{\partial\tau^{2}}\right)_{\delta}\right]
\label{eq:cv_HEOS}
\end{equation}
\begin{equation}
\frac{\tilde{c}_{p}}{R}=\frac{\tilde{c}_{v}}{R}+\frac{\left[1+\delta\left(\frac{\partial\alpha^{r}}{\partial\delta}\right)_{\tau}-\delta\tau\left(\frac{\partial^{2}\alpha^{r}}{\partial\delta\partial\tau}\right)\right]^{2}}{1+2\delta\left(\frac{\partial\alpha^{r}}{\partial\delta}\right)_{\tau}+\delta^{2}\left(\frac{\partial^{2}\alpha^{r}}{\partial\delta^{2}}\right)_{\tau}}
\label{eq:cp_HEOS}
\end{equation}
To complete the mixture model setup, we still need to specify the mixing rules in order to evaluate the reduced mixture density $\delta$, reduced mixture temperature $\tau$ and the departure function $\alpha^E$. We also need to specify the equations of state for $CO_{2}$ and $N_{2}$ that we will use to calculate the ideal Helmholtz energy $\alpha_{i}^{0}$, residual Helmholtz energy $\alpha_{i}^{r}$ and their derivatives. One reason for considering LJ-1999 mixture model and the GERG-2008 mixture model is that they both consider the same set of equations of state for the pure components of $CO_{2}$ and $N_{2}$.
\subsubsection{LJ-1999 Mixture Model} \label{subsubsec:LJ_1999}
As mentioned before, a mixing rule specifies how the equations of state of the pure components will be combined to evaluate the properties of the mixture. Firstly, we require the evaluation of the reduced mixture density and temperature which depend on the expressions of the reducing functions of density and temperature. For the LJ-1999 mixture model, they are given by
 \begin{align}
 \rho_{r} & =  \left[\sum_{i=1}^{n}\frac{x_{i}}{\tilde{\rho}_{ci}}+\sum_{i=1}^{n-1}\sum_{j=i+1}^{n}x_{i}x_{j}\xi_{ij}\right]^{-1}
 \label{eq:rho_red_LJ1999}\\
 T_{r} & =  \sum_{i=1}^{n}x_{i}T_{ci}+\sum_{i=1}^{n-1}\sum_{j=i+1}^{n}x_{i}^{\beta_{ij}}x_{j}^{\phi_{ij}}\zeta_{ij}
 \label{eq:T_red_LJ1999}
 \end{align}
 where $\tilde{\rho}_{ci}$ is the critical amount of substance density of component $i$, $T_{ci}$ is the critical temperature of component $i$, and $\xi_{ij}$, $\beta_{ij}$, $\phi_{ij}$ and $\zeta_{ij}$
 are constant parameters particular to the mixture. For a binary mixture, the expressions for reducing values simplify to 
 \begin{align}
 \rho_{r} & = \left[\frac{x_{1}}{\tilde{\rho}_{c1}}+\frac{x_{2}}{\tilde{\rho}_{c2}}+x_{1}x_{2}\xi_{12}\right]^{-1}\label{eq:rho_red_lemmon-1}\\
 T_{r} & = x_{1}T_{c1}+x_{2}T_{c2}+x_{1}^{\beta_{12}}x_{2}^{\phi_{12}}\zeta_{12}\label{eq:T_red_lemmon-1}
 \end{align}
For the LJ-1999 mixture model, the departure function is given by
\begin{equation}
\frac{A^E}{RT} = \alpha^{E} =  \sum_{i=1}^{n-1}\sum_{j=i+1}^{n}x_{i}x_{j}F_{ij}\sum_{k=1}^{10}N_{k}\delta^{d_{k}}\tau^{t_{k}}
\label{eq:alpha_E_LJ1999}
\end{equation}
The parameters in Eq.\ref{eq:alpha_E_LJ1999} which are not specific to the mixture are presented in the Table \ref{tab:param_alpha_E_LJ1999}. In the case of the binary mixture $CO_{2}-N_{2}$, $F_{12}=2.780647$, $\xi_{12}=0.00659978\, \text{d\ensuremath{m^{3}}mo\ensuremath{l^{-1}}}$, $\zeta_{12}=-31.149300\,\text{K}$, $\phi_{12}=1$ and $\beta_{12}=1$.
\begin{table}[ht]
	\begin{centering}
		\begin{tabular}{|c|l|c|c|}
			\hline 
			$k$ & $N_{k}$ & $d_{k}$ & $t_{k}$\tabularnewline
			\hline 
			\hline 
			1 & $-0.245476271425\times10^{-1}$ & $1$ & $2$\tabularnewline
			\hline 
			2 & $-0.241206117483$ & $1$ & $4$\tabularnewline
			\hline 
			3 & $-0.513801950309\times10^{-2}$ & $1$ & $-2$\tabularnewline
			\hline 
			4 & $-0.239824834123\times10^{-1}$ & 2 & 1\tabularnewline
			\hline 
			5 & $\,\,0.259772344008$ & 3 & 4\tabularnewline
			\hline 
			6 & $-0.172014123104$ & 4 & 4\tabularnewline
			\hline 
			7 & $\,\,0.429490028551\times10^{-1}$ & 5 & 4\tabularnewline
			\hline 
			8 & $-0.202108593862\times10^{-3}$ & 6 & 0\tabularnewline
			\hline 
			9 & $-0.382984234857\times10^{-2}$ & 6 & 4\tabularnewline
			\hline 
			10 & $\,\,0.262992331354\times10^{-5}$ & 8 & -2\tabularnewline
			\hline 
		\end{tabular}
		\par\end{centering}
	
	\protect\caption{Parameters for Eq.\ref{eq:alpha_E_LJ1999} \label{tab:param_alpha_E_LJ1999}}
\end{table}

\subsubsection{GERG-2008 Model} \label{subsubsec:GERG_2008}
The mathematical structure of the reducing functions for density and temperature for the GERG-2008 model are more complicated than the LJ-1999 model and are given by 
\begin{align}
\rho_{r} & =  \left[\sum_{i=1}^{n}x_{i}^2\frac{1}{\tilde{\rho}_{c,i}^2}+\sum_{i=1}^{n-1}\sum_{j=i+1}^{n}2x_{i}x_{j}\beta_{v,ij}\gamma_{v,ij} \cdot \frac{x_i + x_j}{\beta_{v,ij}^{2}x_i + x_j} \cdot \frac{1}{8}\left(\frac{1}{\tilde{\rho}_{c,i}^{1/3} + \tilde{\rho}_{c,j}^{1/3}} \right)^{3} \right]^{-1}\label{eq:rho_red_gerg2008}\\
T_{r} & =  \sum_{i=1}^{n}x_{i}^2 T_{ci} + \sum_{i=1}^{n-1} \sum_{j=i+1}^{n} 2x_{i}x_{j}\beta_{T,ij}\gamma_{T,ij} \cdot \frac{x_i + x_j}{\beta_{T,ij}^{2}x_i + x_j}(T_{c,i}\cdot T_{c,j})^{0.5}\label{eq:T_red_gerg2008}
\end{align}
where $\beta_{v,12}=0.977794634$, $\gamma_{v,12}=1.047578256$, $\beta_{T,12}=1.005894529$ and $\gamma_{T,12}=1.107654104$ for the binary mixture of $CO_{2}-N_{2}$. The function $\alpha^r_{ij}$ which is a part of $\alpha^E$ (Eq.\ref{eq:func_alphaE_HEOS}) is given by
\begin{equation}
	\alpha^r_{12}(\delta, \tau) = \sum_{k=1}^{2} n_{k} \delta^{d_{k}} \tau^{t_{k}} + \sum_{k=3}^{6} n_{k} \delta^{d_{k}} \tau^{t_{k}} \cdot \exp \left[ -\eta_{k} (\delta - \epsilon_{k})^2 - \beta_{k}(\delta - \gamma_{k})\right]
	\label{eq:alpha_r_CO2_N2_GERG2008}
\end{equation}
and $F_{12}=1.0$ for $CO_{2}-N_{2}$. The values of the different parameters in Eq.\ref{eq:alpha_r_CO2_N2_GERG2008} are given in Table \ref{tab:param_alpha_E_GERG2008}.
\begin{table}[ht]
	\begin{centering}
	\begin{tabular}{|c|c|c|c|c|c|c|c|}
		\hline 
		$k$ & $d_k$ & $t_k$ & $n_k$ & $\eta_k$ & $\epsilon_k$ & $\beta_k$ & $\gamma_k$
		\tabularnewline
		\hline 
		\hline 
		1 & 2 & 1.850 & 0.28661625028399 & 0.000 & 0.000 & 0.000 & 0.000\tabularnewline
		\hline 
		2 & 3 & 1.400 & -0.10919833861247 & 0.000 & 0.000 & 0.000 & 0.000\tabularnewline
		\hline 
		3 & 1 & 3.200 & -1.13740320822700 & 0.250 & 0.500 & 0.750 & 0.500\tabularnewline
		\hline 
		4 & 1 & 2.500 & 0.76580544237358 & 0.250 & 0.500 & 1.000 & 0.500\tabularnewline
		\hline 
		5 & 1 & 8.000 & 0.00426380009268 & 0.000 & 0.500 & 2.000 & 0.500\tabularnewline
		\hline 
		6 & 2 & 3.750 & 0.17673538204534 & 0.000 & 0.500 & 3.000 & 0.500\tabularnewline
		\hline 
	\end{tabular}
\par\end{centering}

\protect\caption{Parameters for Eq.\ref{eq:alpha_E_LJ1999} \label{tab:param_alpha_E_GERG2008}}
\end{table}
\subsection{Equation of State for \texorpdfstring{$CO_{2}$}{CO2} and \texorpdfstring{$N_2$}{N2}\label{sub:EOS_CO2_N2}}
For the pure fluids of carbon dioxide $CO_2$ and nitrogen $N_2$, we use the equations of state proposed by \citet{Span1996} and \citet{Span2000} respectively. As mentioned before, the LJ-1999 and GERG-2008 mixture models define mixing rules considering these pure fluid equations of state. They are also accurate in a vast temperature and pressure region as can be seen in Table \ref{tab:refs_consts_pure_CO2_N2_HEOS}. For $CO_2$, the equation of state can be extrapolated from the triple-point temperature down to $90\,\text{K}$ \citep{klimeck2000entwicklung,Kunz2012} without loss in accuracy and we can thus cover the entire range of $p$ and $T$ in the Venus atmosphere.
\begin{table}[ht]
	\begin{centering}
		\footnotesize
		\begin{tabular}{|c|c|c|c|c|c|c|}
			\hline 
			\multirow{2}{*}{Substance} & \multirow{2}{*}{Reference} & \multicolumn{2}{c|}{Range of validity} &  Molar mass & $T_c$ & $\tilde{\rho}_c$ \tabularnewline
			\cline{3-4} 
			&  & $T$ [K]  & Max. $p$ [MPa] & [$\text{kg\ensuremath{\cdot}kmo\ensuremath{l^{-1}}}$] & [K] & [$\text{kmol\ensuremath{\cdot m^{-3}}}$] \tabularnewline
			\hline  
			\hline 
			$CO_2$ & \cite{Span1996} & $216 - 1100$  & $800$ & $44.0098$ & $304.1282$ & $10.6249$ \tabularnewline
			\hline 
			$N_2$ & \cite{Span2000} & $63.151 -1000$ & $2200$ & $28.01348$ & $126.192$ & $11.1839$ \tabularnewline
			\hline 
		\end{tabular}
		\par\end{centering}
	\protect\caption{References and critical parameters of $CO_2$ and $N_2$ 
		\label{tab:refs_consts_pure_CO2_N2_HEOS}}
\end{table}

The equation of state for the pure fluids is explicit in the dimensionless Helmholtz energy $\alpha$ using independent variables of reduced density and temperature.
\begin{equation}
\frac{A_{i}(\rho,T)}{RT}=\alpha_{i}(\delta,\tau)=\alpha_{i}^{o}(\delta,\tau)+\alpha_{i}^{r}(\delta,\tau)\label{eq:eos_pure_fluid}
\end{equation}
where the subscript $i$ denotes the component of interest (i.e. $CO_2$ or $N_2$). In the above equation, $\delta$ is the mixture reduced density and $\tau$ is the mixture reduced temperature when calculating the contribution of component $i$ to any mixture. When calculating the Helmholtz energy for a system containing only the pure fluid $i$, $\delta=\tilde{\rho}/\tilde{\rho}_{c}$ and $\tau=T_{c}/T$. This would also be obtained from the reducing functions of Eqs.\ref{eq:rho_red_LJ1999} and \ref{eq:T_red_LJ1999} for the LJ-1999 mixture model or Eqs.\ref{eq:T_red_gerg2008} and \ref{eq:rho_red_gerg2008} for the GERG-2008 mixture model respectively 
\paragraph{Nitrogen}The ideal gas Helmholtz energy of $N_{2}$ is given by
\begin{equation}
\alpha_{N_{2}}^{o}(\delta,\tau)=\ln\delta+a_{1}\ln\tau+a_{2}+a_{3}\tau+a_{4}\tau^{-1}a_{5}\tau^{-2}+a_{6}\tau^{-3}+a_{7}\ln[1-exp(-a_{8}\tau)]\label{eq:alpha_0_N2}
\end{equation}
where $a_{1}=2.5$, $a_{2}=-12.76953$, $a_{3}=-0.007841630$, $a_{4}=-1.934819\times10^{-4}$,
$a_{5}=-1.247742\times10^{-5}$, $a_{6}=6.678326\times10^{-8}$, $a_{7}=1.012941$
and $a_{6}=26.65788$.
The residual gas Helmholtz energy of $N_{2}$ is given by 
\begin{dmath}
\alpha_{N_{2}}^{r}(\delta,\tau)=\sum_{k=1}^{6}N_{k}\delta^{i_{k}}\tau^{j_{k}}+\sum_{k=7}^{32}N_{k}\delta^{i_{k}}\tau^{j_{k}}\exp(-\delta^{l_{k}})+\sum_{k=33}^{36}N_{k}\delta^{i_{k}}\tau^{j_{k}}\exp(-\psi_{k}(\delta-1)^{2}-\beta_{k}(\tau-\gamma_{k})^{2})\label{eq:alpha_r_N2}
\end{dmath}
The derivatives of $\alpha_{N_2}^r$ as required in the mixture model and values of the parameters $N_k$, $i_k$, $j_k$, $l_k$, $\psi_k$, $\beta_k$, and $\gamma_k$ (for different values of $k$) are given in \ref{appendix_eos_N2}.
\paragraph{Carbon Dioxide} Ideal Helmholtz energy is given by
\begin{equation}
\alpha_{CO_{2}}^{o}(\delta,\tau)=\ln\delta+a_{1}^{0}+a_{2}^{0}\tau+a_{3}^{0}\ln\tau+\sum_{i=4}^{8}a_{i}^{0}\ln[1-\exp(-\tau\theta_{i}^{0})]\label{eq:alpha_0_CO2}
\end{equation}
\begin{table}[H]
	\begin{centering}
		\begin{tabular}{ccr||ccr}
			\hline 
			$i$ & $a_{i}^{0}$ & $\theta_{i}^{0}$ & $i$ & $a_{i}^{0}$ & $\theta_{i}^{0}$\tabularnewline
			\hline 
			1 & 8.37304456 &  & 5 & 0.62105248 & 6.11190\tabularnewline
			2 & -3.70454304 &  & 6 & 0.41195293 & 6.77708\tabularnewline
			3 & 2.50000000 &  & 7 & 1.04028922 & 11.32384\tabularnewline
			4 & 1.99427042 & 3.15163 & 8 & 0.08327678 & 27.08792\tabularnewline
		\end{tabular}
		\par\end{centering}
	
	\protect\caption{Parameters as in Eq.\ref{eq:alpha_0_CO2}}
\end{table}
Residual Helmholtz energy is given by
\begin{dmath}
	\alpha_{CO_{2}}^{r}(\delta,\tau) = \sum_{i=1}^{7}n_{i}\delta^{d_{i}}\tau^{t_{i}}+\sum_{i=8}^{34}n_{i}\delta^{d_{i}}\tau^{t_{i}}\exp(-\delta^{c_{i}})+\sum_{i=35}^{39}n_{i}\delta^{d_{i}}\tau^{t_{i}}\exp(-\alpha_{i}(\delta-\epsilon_{i})^{2}-\beta_{i}(\tau-\gamma_{i})^{2}) +\sum_{i=40}^{42}n_{i}\Delta^{b_{i}}\delta\Psi\label{eq:alpha_r_CO2}
\end{dmath}
with
\begin{eqnarray}
\theta & = & (1-\tau)+A_{i}[(\delta-1)^{2}]^{1/(2\beta_{i})}\\
\Delta & = & \theta^{2}+B_{i}[(\delta-1)^{2}]^{a_{i}}\\
\Psi & = & \exp(-C_{i}(\delta-1)^{2}-D_{i}(\tau-1)^{2})
\end{eqnarray}
The derivatives of $\alpha_{CO_2}^r$, $\theta$, $\Delta$ and $\Psi$ as required in the mixture model and values of the parameters $n_i$, $d_i$, $t_i$, $c_i$, $\alpha_i$, $\beta_i$, $\gamma_i$, $\epsilon_i$, $a_i$, $b_i$, $A_i$, $B_i$, $C_i$ and $D_i$ are given in \ref{appendix_eos_CO2}.
\subsection{Ideal Mixture of Real Gases Model} \label{subsec:IMRG}
We will now discuss how the approach in \citep{Seiff1980} can be followed using the equations of state in Helmholtz energy. To obtain the thermodynamic properties of ideal mixture of real gases (IMRG), the first step is to neglect the contribution of non-ideal interactions between the different components in the mixture model. This can be done by setting $\alpha^E$ to zero. The non-dimensionalized Helmholtz free energy of the IMRG can then be written as
 \begin{equation}
 \alpha^{IMRG}(\tilde{\rho},T,\bar{x}) = \sum \limits_{i=1}^{n}x_{i}\left[\alpha_{i}^{o}(\tilde{\rho},T)+\ln x_{i} + \alpha_{i}^{r}(\delta_i,\tau_i)\right]
 \end{equation}
where $\delta_i = \tilde{\rho}/\tilde{\rho}_{c,i}$ is the reduced density and $\tau_i = T_{c,i}/T$ is the reduced temperature of component $i$. It is important to note that $\alpha^r_i$ is not a function of the mixture reduced density $\delta$ and mixture reduced temperature $\tau$ here. These reduced values depend on mixing rules which vary from one real gas mixture model to another as we have seen in the case of LJ-1999 and GERG-2008 models. The IMRG must not depend on the mixing rule being used. This approach is similar to that followed in \citep{asmestandards2012}.
The thermodynamic properties of the IMRG can be obtained by using Gibbs-Dalton law which is valid for ideal mixtures
\begin{equation}
p = \sum \limits_{i=1}^n  x_i p_i ;\quad \tilde{c}_v = \sum \limits_{i=1}^n x_i \tilde{c}_{v,i} ;\quad \tilde{c}_p = \sum \limits_{i=1}^n x_i \tilde{c}_{p,i}
\end{equation}
where $p_i$ is the partial pressure of component $i$ for $\tilde{\rho}$ and $T$ which can be evaluated using Eq.\ref{eq:p_HEOS}. Similarly, $\tilde{c}_{v,i}$ is the partial specific isochoric heat capacity and $\tilde{c}_{p,i}$ is the partial isobaric specific heat capacity of component $i$ which can be evaluated using Eqs.\ref{eq:cv_HEOS} and \ref{eq:cp_HEOS}. Specifically, we have
\begin{equation}
p= \sum \limits_{i=1}^n \tilde{\rho}RT x_i \left[1+ \delta_i \left(\frac{\partial\alpha^{r}_i}{\partial\delta_i}\right)_{\tau_i}\right]
\label{eq:p_IMRG_HEOS}
\end{equation}
\begin{equation}
\frac{\tilde{c}_{v}}{R}= - \sum \limits_{i=1}^n x_i \tau^{2}_i \left[\left(\frac{\partial^{2}\alpha^{0}_i}{\partial\tau^{2}_i}\right) + \left(\frac{\partial^{2}\alpha^{r}_i}{\partial\tau^{2}_i}\right)_{\delta_i}\right]
\label{eq:cv_IMRG_HEOS}
\end{equation}
\begin{equation}
\frac{\tilde{c}_{p}}{R}=\frac{\tilde{c}_{v}}{R} + \sum \limits_{i=1}^n x_i \frac{\left[1 + \delta_i\left(\frac{\partial\alpha^{r}_i}{\partial\delta_i}\right)_{\tau_i} - \delta_i \tau_i \left(\frac{\partial^{2}\alpha^{r}_i}{\partial\delta_i\partial\tau_i}\right)\right]^{2}}{1 + 2\delta_i \left(\frac{\partial\alpha^{r}_i}{\partial\delta_i}\right)_{\tau_i}+\delta^{2}_i\left(\frac{\partial^{2}\alpha^{r}_i}{\partial\delta^{2}_i}\right)_{\tau_{i}}}
\label{eq:cp_IMRG_HEOS}
\end{equation}
\section{Density Solvers} \label{sec:Density_Solvers}
As we have seen in the previous section, the independent variables for the mixture models in Helmholtz free energy are amount of substance density $\tilde{\rho}$ and temperature $T$. When pressure $p$ and temperature are available to us, we need to solve for amount of substance density in Eqs.\ref{eq:p_HEOS} and \ref{eq:p_IMRG_HEOS}. We use MATLAB's inbuilt function fzero for root finding which uses a combination of bisection, secant, and inverse quadratic interpolation methods. The equations (Eqs.\ref{eq:p_HEOS} and \ref{eq:p_IMRG_HEOS}) for which we need to obtain roots are highly nonlinear and many roots are possible. It is thus important to ascertain which root $\tilde{\rho}$ is physically meaningful. We follow the suggestions in \citep{gernert2014calculation} to do this. MATLAB's root finding solver fzero requires an initial estimate or interval for $\tilde{\rho}$ in which we believe the root lies in. This was generated using the ideal gas law or exploration of the range of Eqs.\ref{eq:p_HEOS} and \ref{eq:p_IMRG_HEOS} for different values of $p$ and $T$. This could also be done through using an SRK equation of state as suggested in \citep{gernert2014calculation}.
\section{Verification of the results with Available Experimental Results\label{sec:Verification-against-Experiments}}
We compare the approaches of pure $CO_2$ model \citep{Staley1970}, IMRG model \citep{Seiff1980}, LJ-1999 model \citep{Lemmon1999} and GERG-2008 model \citep{Kunz2012} against experimental data. Our main motivation is to show that the real gas mixture models perform better than the other approaches in predicting the thermodynamic properties of the real gas mixture $CO_{2}-N_{2}$. \citet{Staley1970} and \citet{Seiff1980} used the compilation of experimentally determined properties of $CO_{2}$ and $N_{2}$ found in \citet{Hilsenrath1955}. To account for more recent experiments, we use the equations of state for $CO_{2}$ \citep{Span1996} and $N_{2}$ \citep{Span2000}. For comparison, the uncertainty in the isobaric specific heat capacity data of $CO_{2}$ tabulated in \citep{Hilsenrath1955} is of the order of $\pm2.0\%$ for $220\,\text{K\,}\leq T\leq\,600\,\text{K}$ at atmospheric pressure. This is considering the experimental data available at that time which had low reliability. The uncertainty in $c_{p}$ for $CO_{2}$ as obtained from the equation of state in Helmholtz energy \citep{Span1996} is of the order of $\pm0.15\%$ at the same pressure when considered against more reliable experimental data. Considering $N_{2}$, the uncertainty in $c_{p}$ data obtained using the equation of state in Helmholtz energy \citep{Span2000} is of the order of $\pm0.3\%$ against that of $\pm3.0\%$ uncertainty in $c_{p}$ data of \citet{Hilsenrath1955} for $100\,\text{K\,}\leq T\leq\,700\,\text{K}$ at atmospheric pressure.

For the comparing accuracy of the different models, we look at experiments that reported results for $CO_{2}-N_{2}$ mixtures with $x_{CO_{2}}>0.9$ as the main contention by the approach proposed by \citet{Seiff1980} was that non-ideal interactions between $CO_{2}$ and $N_{2}$ can be safely neglected for such mixtures. Table \ref{tab:Expt_CO2_N2} summarizes the literature that was used.
\begin{table}[H]
\begin{centering}
\begin{tabular}{|r|c|c|c|c|}
\hline 
{\footnotesize{}Reference} & {\footnotesize{}Type of Expt. Data} & {\footnotesize{}Pressure Range (MPa)} & {\footnotesize{}Temperature Range (K)} & {\footnotesize{}$x_{CO_{2}}$ (\%)}\tabularnewline
\hline 
\hline 
{\footnotesize{}\citet{Brugge1989}} & {\footnotesize{}$p\rho T$} & {\footnotesize{}0.21-6.63} & {\footnotesize{}300-320} & {\footnotesize{}90.92}\tabularnewline
\hline 
{\footnotesize{}\citet{Brugge1997}} & {\footnotesize{}$p\rho T$} & {\footnotesize{}1.03-69.09} & {\footnotesize{}285-450} & {\footnotesize{}90.92}\tabularnewline
\hline 
{\footnotesize{}\citet{Ely1989}} & {\footnotesize{}$p\rho T$} & {\footnotesize{}2.26-33.10} & {\footnotesize{}250-330} & {\footnotesize{}98.20}\tabularnewline
\hline 
{\footnotesize{}\citet{Mantovani2012}} & {\footnotesize{}$p\rho T$} & {\footnotesize{}1.00-20.00} & {\footnotesize{}303-383} & {\footnotesize{}90.21, 95.85}\tabularnewline
\hline 
{\footnotesize{}\citet{Bishnoi1972}} & {\footnotesize{}$c_{p}$} & {\footnotesize{}3.45-14.48} & {\footnotesize{}313-363} & {\footnotesize{}93.23}\tabularnewline
\hline 
\end{tabular}
\par\end{centering}

\protect\caption{Experimental Data for $CO_{2}$ Rich Mixtures of $CO_{2}-N_{2}$\label{tab:Expt_CO2_N2}}
\end{table}

\begin{figure}[H]
	\centering
	\begin{subfigure}[t]{0.5\textwidth}
		\centering
		\includegraphics[scale=0.55]{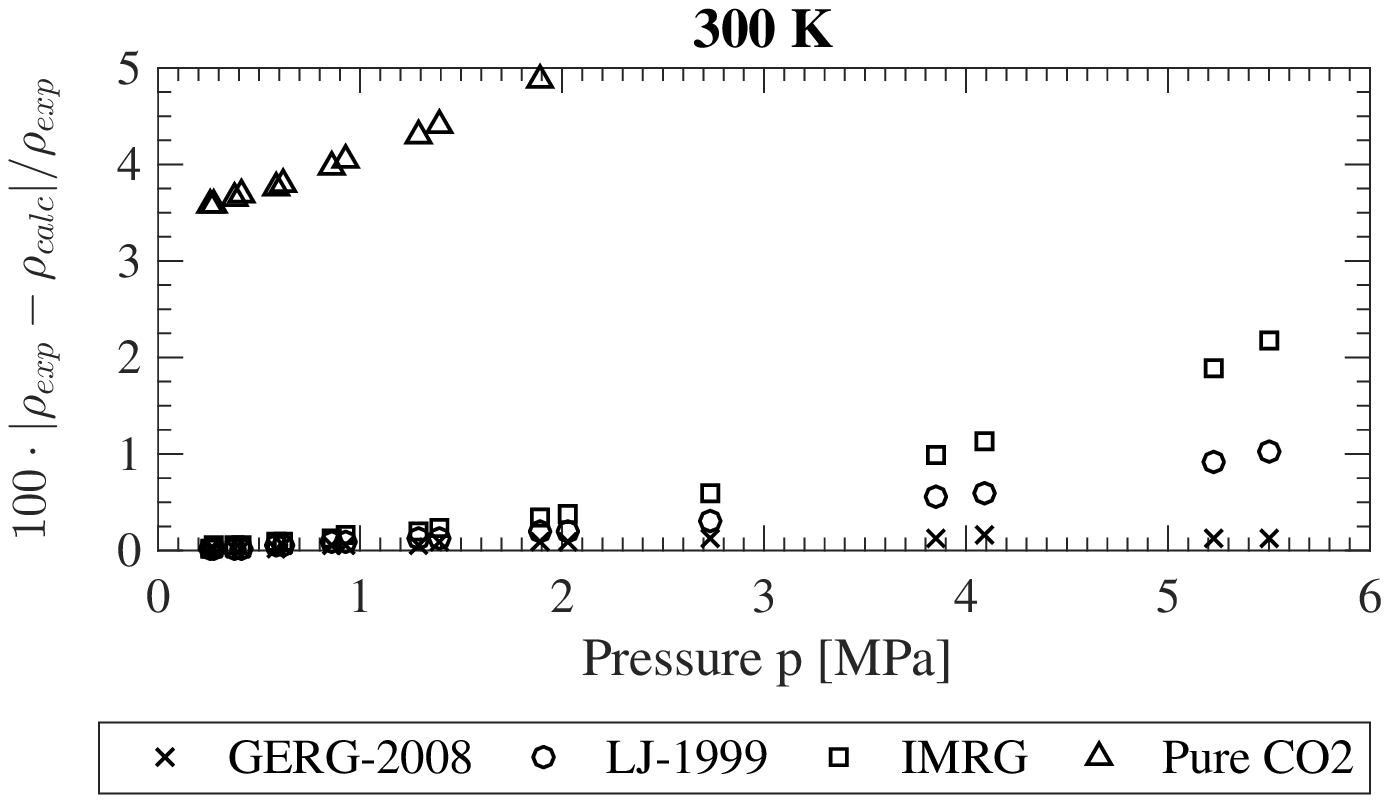}
	\end{subfigure}%
	~ 
	\begin{subfigure}[t]{0.5\textwidth}
		\centering
		\includegraphics[scale=0.55]{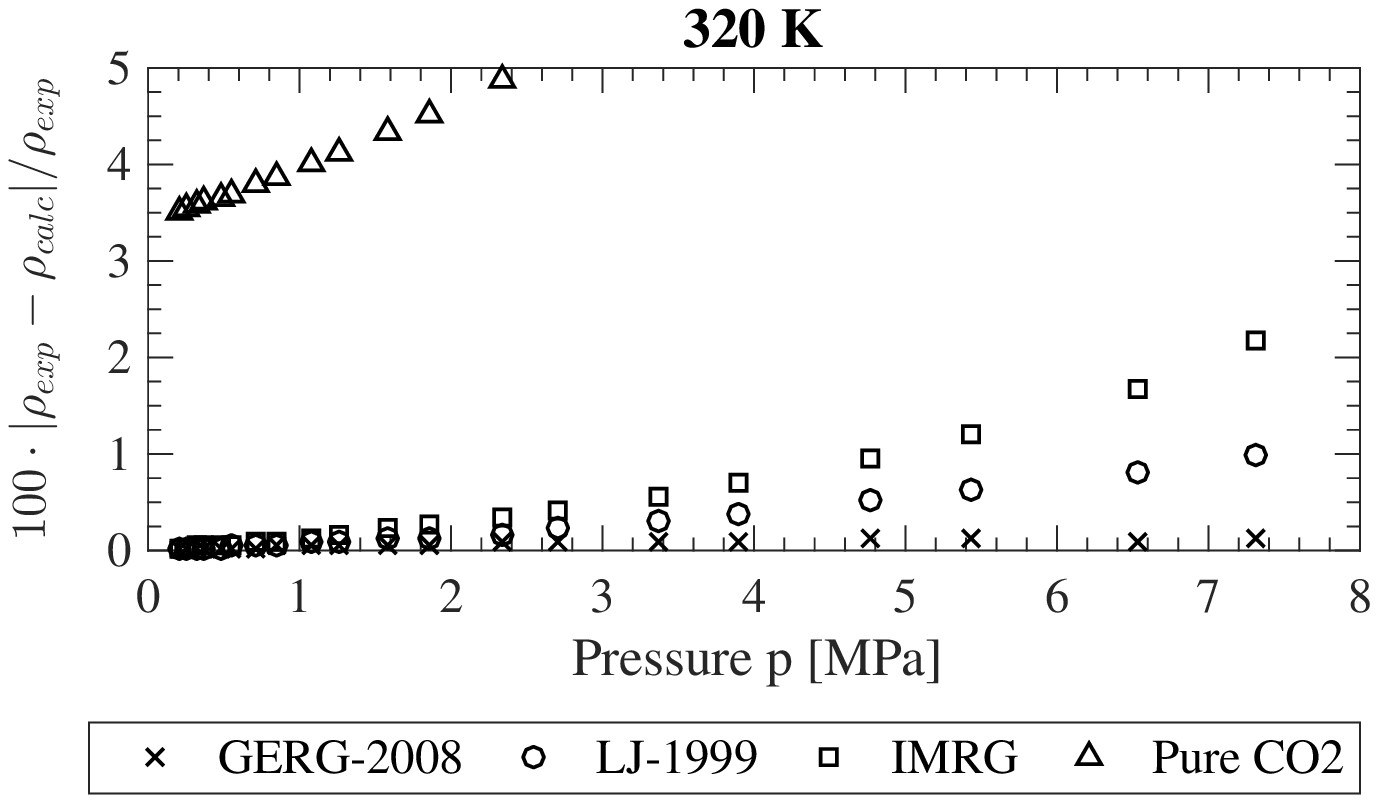}
	\end{subfigure}
	\caption{Deviations of density calculated using the different models from experimental data for $x_{CO_2}=0.90921$ in \citep{Brugge1989}}
	\label{fig:Expt_Comp_Brugge1989}
\end{figure}

\begin{figure}[H]
	\centering
	\begin{subfigure}[t]{0.5\textwidth}
		\centering
		\includegraphics[scale=0.55]{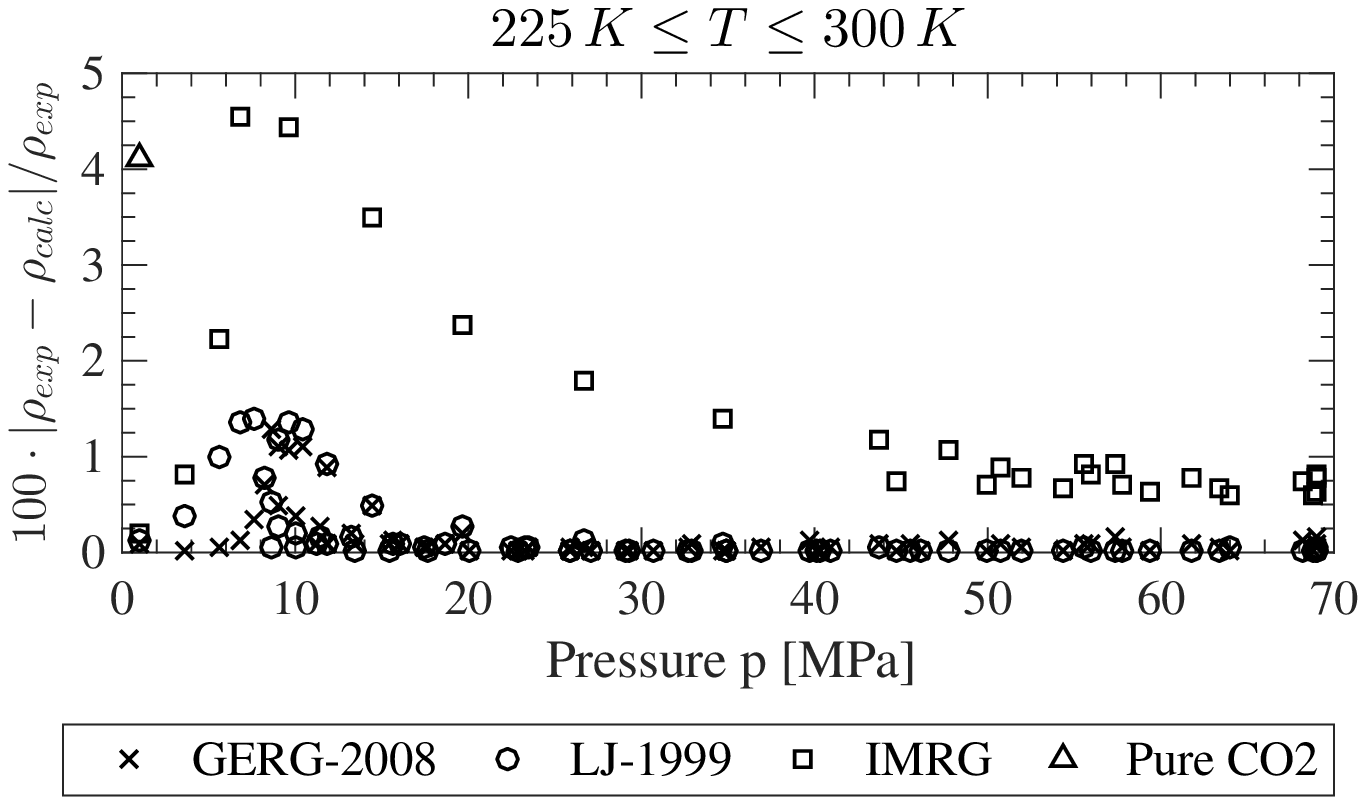}
	\end{subfigure}%
	~ 
	\begin{subfigure}[t]{0.5\textwidth}
		\centering
		\includegraphics[scale=0.55]{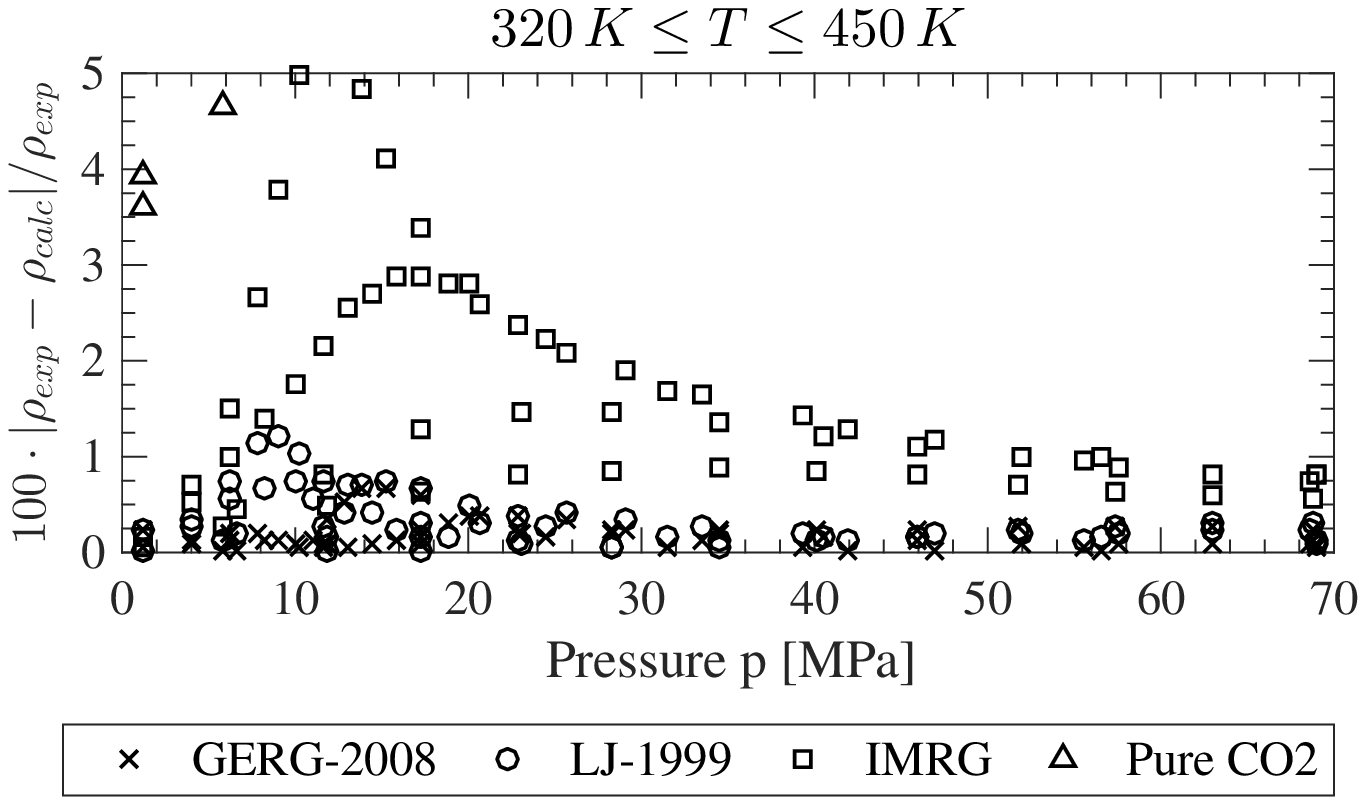}
	\end{subfigure}
	\caption{Deviations of density calculated using the different models from experimental data for $x_{CO_2}=0.90921$ in \citep{Brugge1997}}
	\label{fig:Expt_Comp_Brugge1997}
\end{figure}

\begin{figure}[H]
	\centering
	\begin{subfigure}[t]{0.5\textwidth}
		\centering
		\includegraphics[scale=0.55]{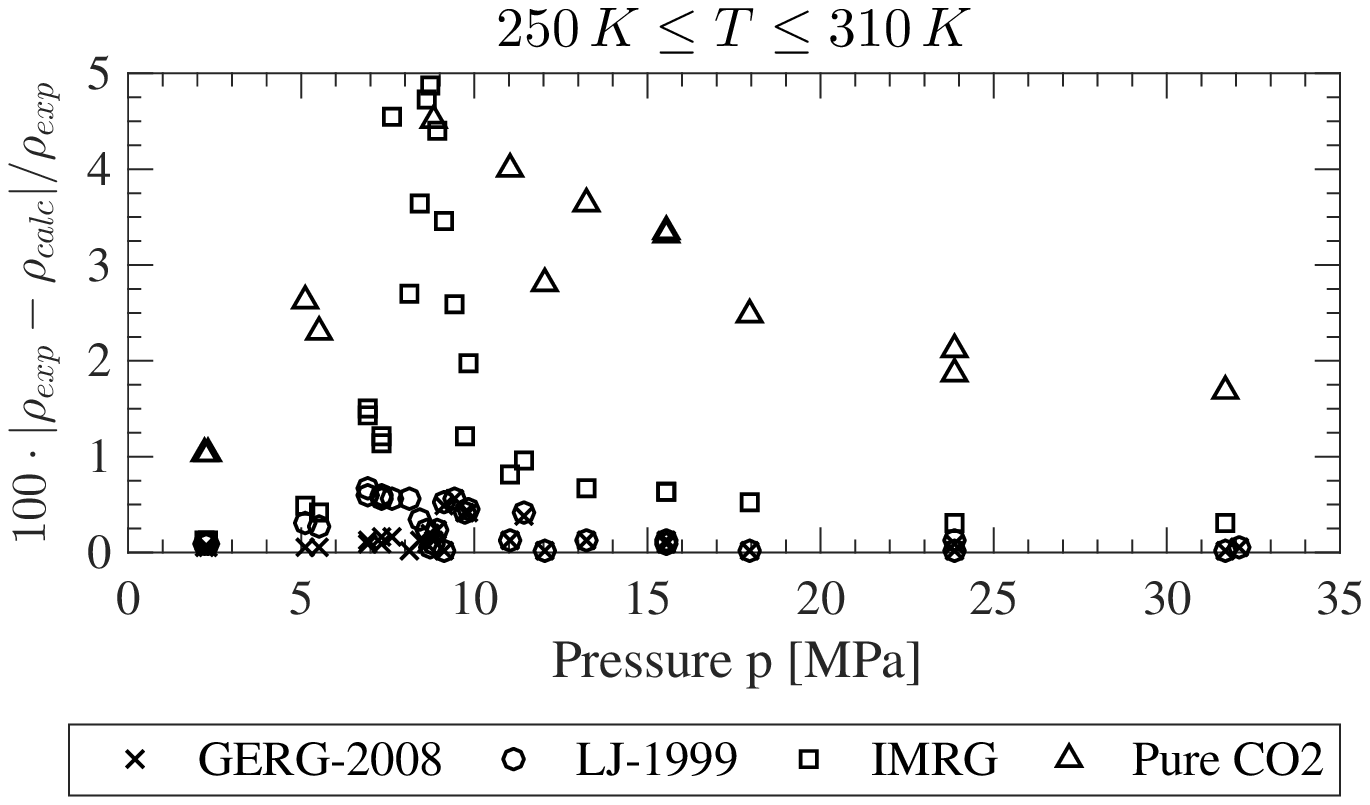}
	\end{subfigure}%
	~ 
	\begin{subfigure}[t]{0.5\textwidth}
		\centering
		\includegraphics[scale=0.55]{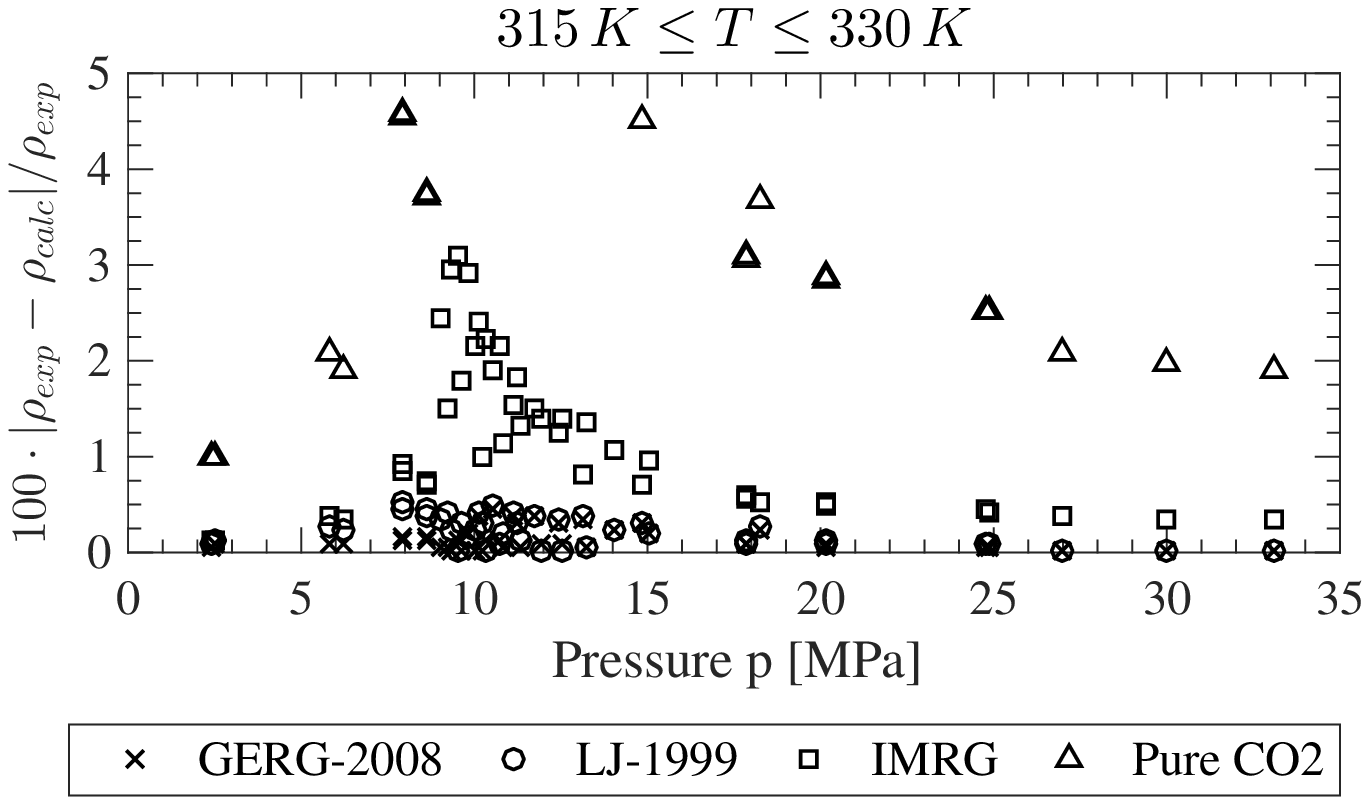}
	\end{subfigure}
	\caption{Deviations of density calculated using the different models from experimental data for $x_{CO_2}=0.982$ in \citep{Ely1989}}
	\label{fig:Expt_Comp_Ely1989}
\end{figure}

\begin{figure}[H]
	\centering
	\begin{subfigure}[t]{0.5\textwidth}
		\centering
		\includegraphics[scale=0.6]{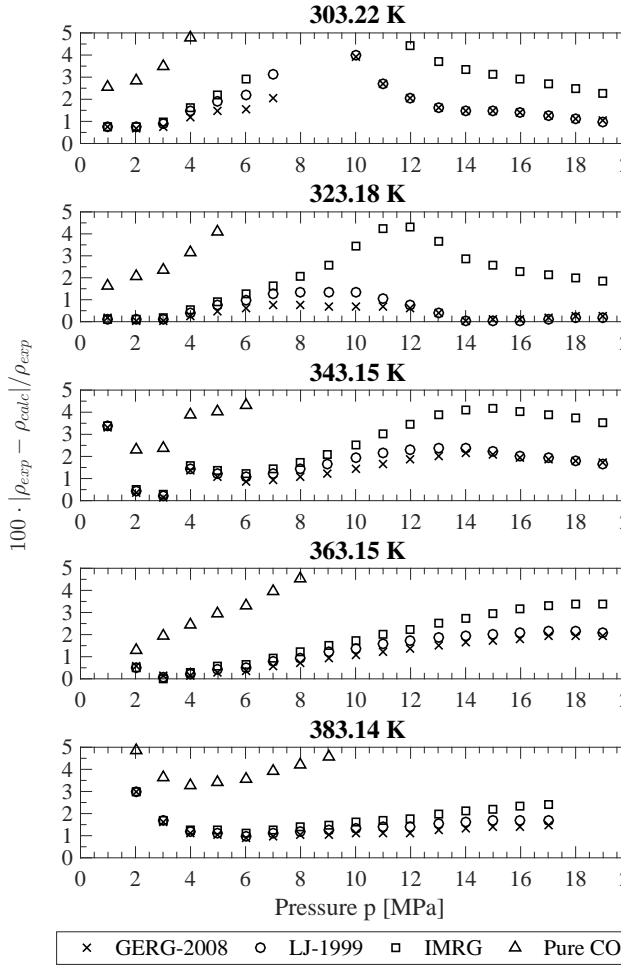}
		\caption{$x_{CO_2}=0.9585$}
	\end{subfigure}%
	~ 
	\begin{subfigure}[t]{0.5\textwidth}
		\centering
		\includegraphics[scale=0.6]{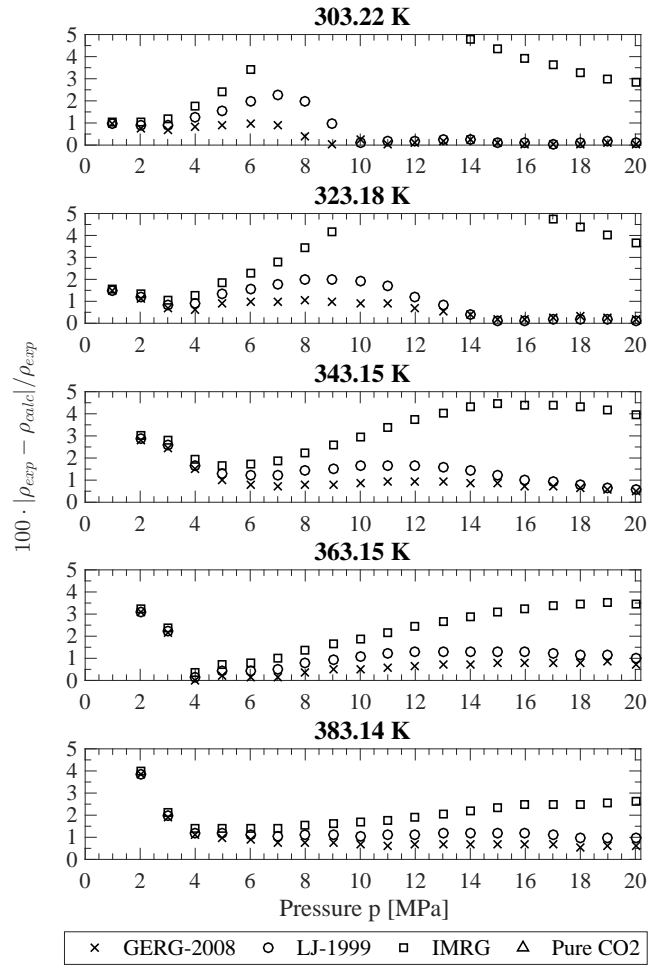}
		\caption{$x_{CO_2}=0.9021$}
	\end{subfigure}
	\caption{Deviations of density calculated using the different models from experimental data in \citep{Mantovani2012}}
	\label{fig:Expt_Comp_Mantovani2012}
\end{figure}

In Figures \ref{fig:Expt_Comp_Brugge1989}, \ref{fig:Expt_Comp_Brugge1997}, \ref{fig:Expt_Comp_Ely1989} and \ref{fig:Expt_Comp_Mantovani2012}, we look at the relative deviations of density calculated using the different models against the experimental data. The results indicate that the GERG-2008 mixture model is the most accurate for these temperature and pressure ranges followed by the LJ-1999 mixture model, then the IMRG model and lastly considering a pure $CO_2$ equation of state. In addition to comparing the trends of deviations, we can compare the percentage average absolute deviations in density (calculated over $N$ data points) which is given by
\begin{equation}
	\text{AAD\%}_{calc - exp} = \frac{1}{N} \sum \limits_{i=1}^{N} 100 \frac{|\rho_{exp} - \rho_{calc}|}{\rho_{exp}}
\end{equation} 
where $\rho_{exp}$ is the experimentally measured value of density and $\rho_{calc}$ is that predicted by the mixture model. For example, the $\text{AAD}\%$ in density obtained from the different mixture models against the experimental data of \citep{Brugge1989} are: (i) GERG-2008 -- 0.0671, (ii) LJ-1999 -- 0.2446, (iii) IMRG -- 0.4842, and (iv) Pure $CO_2$ -- 5.7316. The real gas mixture models are also able to give accurate values of density for the $CO_2 - N_2$ mixture in the supercritical region. Considering $x_{CO_2}=0.9585$, the $\text{AAD}\%$ in density obtained from the different mixture models against the experimental data of \citep{Mantovani2012} are: (i) GERG-2008 -- 1.3592, (ii) LJ-1999 -- 1.5421, (iii) IMRG -- 2.6761, and (iv) Pure $CO_2$ -- 9.8766.

\begin{figure}[H]
	\centering
	\includegraphics[scale=0.8]{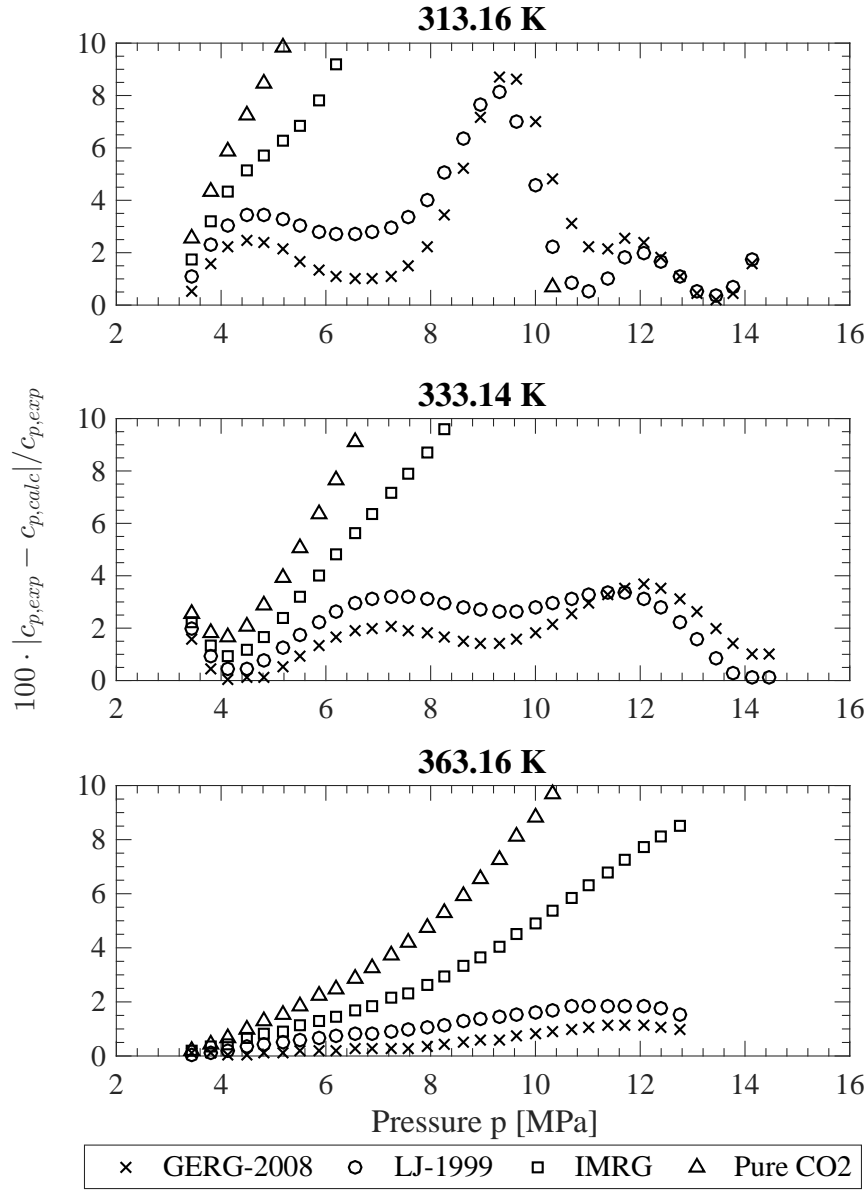}
	\caption{Deviations of isobaric heat capacities calculated using the different models from experimental data for $x_{CO_2}=0.9323$ in \citep{Bishnoi1972}}
	\label{fig:Expt_Comp_Bishnoi1972}
\end{figure}

This indicates that the real gas mixture models can be used with confidence in calculating accurate values of the thermodynamic properties for the $CO_2 - N_2$ mixture which exists in a supercritical state in the lower parts of the Venus atmosphere.
Lastly, we look at the relative deviations of isobaric specific heat capacity calculated using the different models against the experimental data \citep{Bishnoi1972}.  The trends in Figure \ref{fig:Expt_Comp_Bishnoi1972} show that the GERG-2008 and LJ-1999 mixture models are far more accurate than the IMRG model and the pure $CO_2$ equation of state at predicting values of $c_p$. The $\text{AAD}\%$ in $c_p$ over the real gas mixture models against the experimental data of \citep{Bishnoi1972} are (i) GERG-2008 -- 1.7083, and (ii) LJ-1999 -- 2.1151.

Through the comparison of the different mixture models against experimental data, we have seen that it is imperative to include the non-ideal interactions of $CO_2$ and $N_2$ in the mixture when calculating the thermodynamic properties of the mixture. Moreover, this served as a verification of our implementation of the different real gas mixture models. The trends of deviations in $\rho$ and $c_p$ obtained here closely match with those in \citep{gernert2013new} for the GERG-2008 model and \citep{Lemmon1996} for the LJ-1999 model for sets of common experimental data.

\section{Adiabatic Lapse Rate\label{sec:Adiabatic-Lapse-Rate}}
\subsection{LJ-1999, GERG-2008 and Pure $CO_2$ Models}
Our starting point for calculating adiabatic lapse rate is Eq.\ref{eq:adiabatic_lapse_rate}
\begin{equation}
\Gamma=-\frac{T}{\rho}\frac{\left(\frac{\partial p}{\partial T}\right)_{\rho}}{\left(\frac{\partial p}{\partial\rho}\right)_{T}}\left(\frac{g}{c_{p}}\right) 
\label{eq:adiabatic_lapse_rate2}
\end{equation}
Isobaric heat capacity $c_p$ can be computed using Eq.\ref{eq:cp_HEOS}. We further note that from Eq.\ref{eq:p_HEOS} and using the definitions of reducing functions, the different partial derivatives of pressure can be computed from
\begin{align}
\left(\frac{\partial p}{\partial T}\right)_{\rho} &= \tilde{\rho} R \left(1 + \delta \alpha^r_{\delta} - \delta \tau \alpha^r_{\delta\tau} \right) \label{eq:dp_dT_HEOS} \\
\left(\frac{\partial p}{\partial\rho}\right)_{T} &= \frac{RT}{M} \left(1 + 2\delta \alpha^r_{\delta} + \delta^2  \alpha^r_{\delta\delta} \right)
\label{eq:dp_drho_HEOS}
\end{align}
The adiabatic lapse rate can then be computed using
\begin{equation}
\Gamma=-\frac{\left(1 + \delta \alpha^r_{\delta} - \delta \tau \alpha^r_{\delta\tau} \right)}{\left(1 + 2\delta \alpha^r_{\delta} + \delta^2  \alpha^r_{\delta\delta} \right)} \left(\frac{g}{c_{p}}\right)
\label{eq:adiabatic_lapse_rate_HEOS}
\end{equation}
In the above expression, acceleration due to gravity was assumed to change only with altitude $z$ as $g=g_{o}\frac{R_{o}^{2}}{(R_{o}+z)^{2}}$ where $g_{o}=8.869m/s^{2}$ and the radius of the planet of Venus $R_{o}$ was considered to be $6052\,\text{km}$. 

The atmospheric conditions of Venus are recorded in terms of pressure $p$ and temperature $T$. As a part of calculating $\Gamma$, density $\rho$ needs to be determined. We follow the discussion in Sec.\ref{sec:Density_Solvers} and additionally consider the initial estimate of density from interpolated values of density reported in \citep{moroz1981atmosphere} for altitude range of $0-100 \text{km}$. 

\subsection{IMRG Model}
We follow the same approach as discussed in \cite{Seiff1980} to calculate the adiabatic lapse rate for the ideal mixture of real gases model (IMRG). The only difference is that the thermodynamic properties of the IMRG model are determined from equations of state in Helmholtz free energy as was discussed in Sec. \ref{subsec:IMRG}. The expression in Eq.\ref{eq:adiabatic_lapse_rate_Seiff} can be calculated using
\begin{equation}
\Gamma=b\left(\frac{g}{c_{p}}\right)
\end{equation}
where $c_{p}$ is calculated using Eq.\ref{eq:cp_IMRG_HEOS} and $b=-aT$ (with $a$ as defined in Eq.\ref{eq:a_Seiff}). For the IMRG model, $b$ can be calculated as
\begin{align}
b & = -\sum_{i}x_{i} \left[ \frac{T}{\rho} \left(\frac{\partial \rho}{\partial T} \right)_{p} \right]_{i}\\
 & = -\sum_{i}x_{i} \left[ \frac{T}{\rho}\frac{\left(\frac{\partial p}{\partial T}\right)_{\rho}}{\left(\frac{\partial p}{\partial\rho}\right)_{T}} \right]_{i}
\end{align}
The expressions for the partial derivatives of $p$ for component $i$ can be written in a similar fashion to those in Eqs.\ref{eq:dp_dT_HEOS} and \ref{eq:dp_drho_HEOS}. The expression for adiabatic lapse rate for the IMRG model then becomes
\begin{equation}
\Gamma=- \left( \sum_{i}x_{i} \left[ \frac{\left(1 + \delta_i \alpha^r_{\delta_i} - \delta_i \tau_i \alpha^r_{\delta_i \tau_i} \right)}{\left(1 + 2\delta_i \alpha^r_{\delta_i} + \delta^2_i  \alpha^r_{\delta_i \delta_i} \right)} \right] \right) \left(\frac{g}{c_{p}}\right)
\label{eq:adiabatic_lapse_rate_IMRG}
\end{equation}
where $\alpha^r$ is different for each component $i$. 

\section{Results and Discussion \label{sec:Results}}
\cite{oyama1980pioneer} reported a vertical gradient of $N_2$ between $22$ and $52$km altitudes. However, for the sake of comparison, we consider the Venus atmosphere to be composed of a real gas binary mixture of $CO_{2}-N_{2}$ in the constant volume mixing ratio of $96.5:3.5$ \citep{VonZahn1983}. Additionally, we can assume within experimental uncertainty that the atmosphere can be considered to be composed of a real gas binary mixture of $CO_{2}-N_{2}$ in a ratio of $96.5:3.5$ by mole fraction. Thus, we neglect any vertical variation in the composition. 

\begin{figure}[H]
	\centering
	\begin{subfigure}[t]{0.5\textwidth}
		\centering
		\includegraphics[scale=0.4]{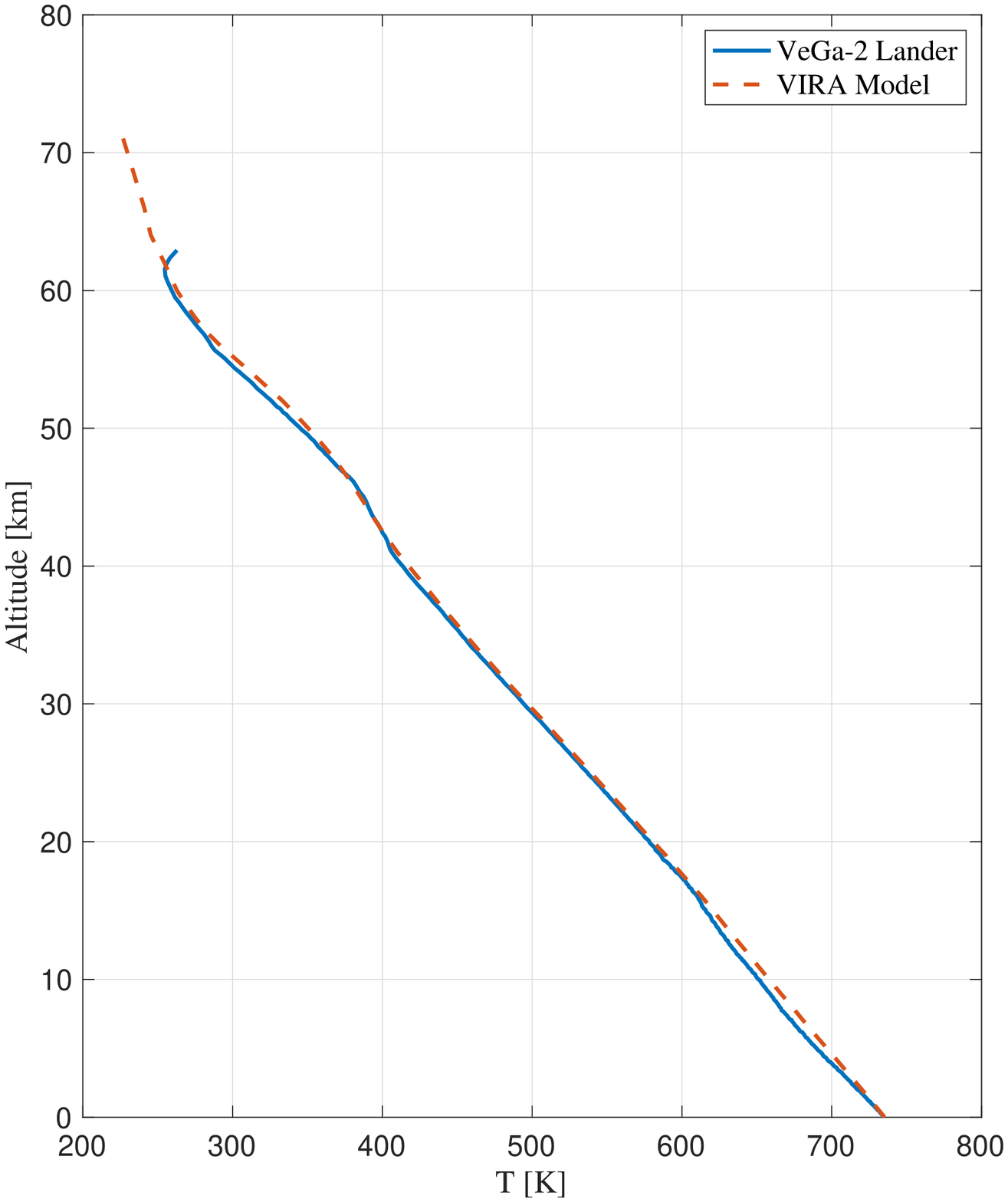}
		\vspace{-1cm}
		\caption{Profiles of temperature with altitude}
		\label{fig:temperature_Seiff1985_Linkin1987}
	\end{subfigure}%
	~
	\begin{subfigure}[t]{0.5\textwidth}
		\centering
		\includegraphics[scale=0.4]{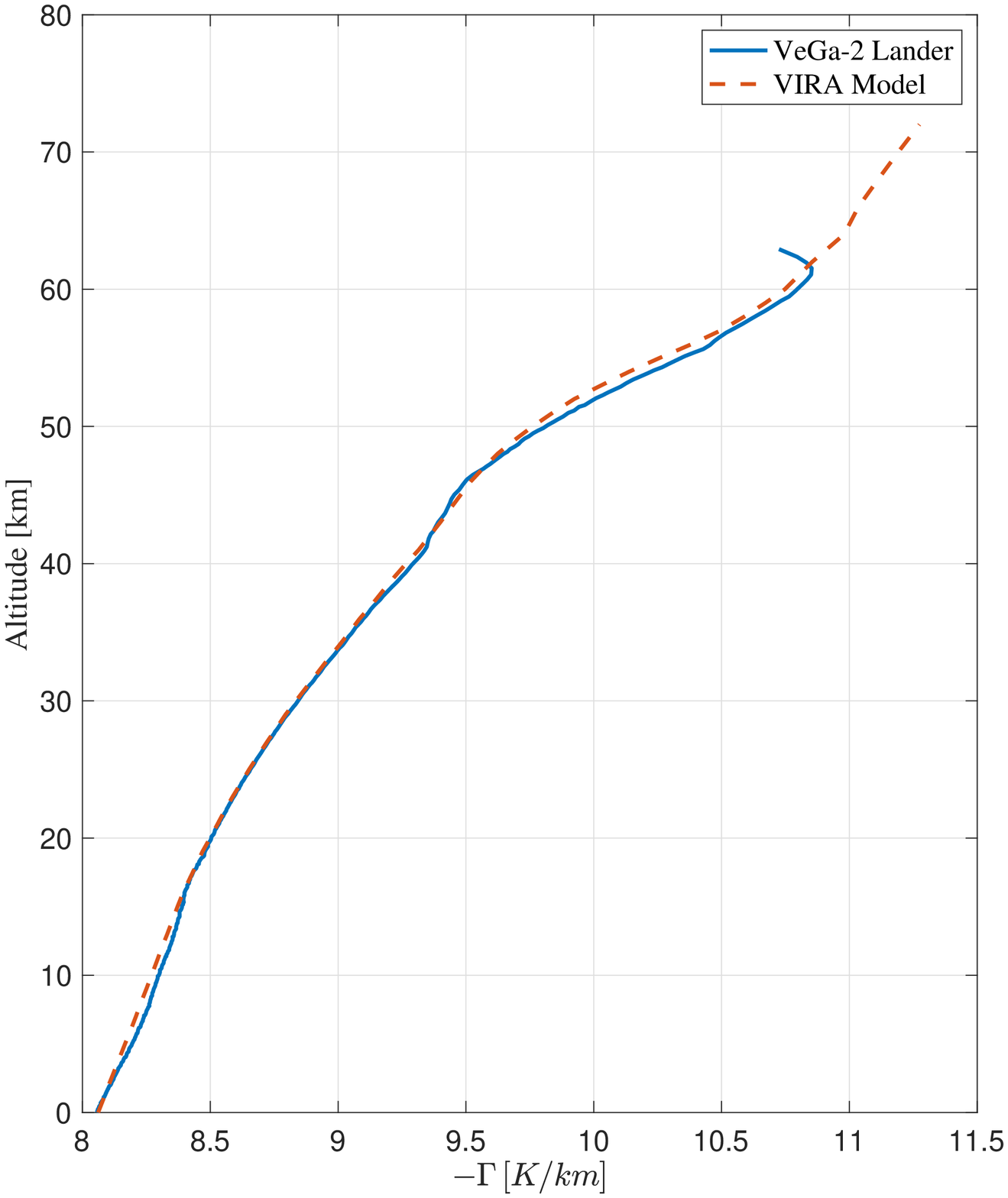}
		\vspace{-1cm}
		\caption{Profiles of adiabatic lapse rate with altitude}
		\label{fig:lapse_rate_Seiff1985_Linkin1987}
	\end{subfigure}
	\caption{Comparison of profiles of temperature and adiabatic lapse rate computed using the GERG-2008 mixture model for the VeGa-2 Lander \citep{Linkin1987} and the VIRA model \citep{Seiff1985} constructed from the four Pioneer Venus probes' data \citep{Seiff1980}}
	\label{fig:profiles_Seiff1985_Linkin1987}
\end{figure}

\begin{figure}[H]
	\centering
	\includegraphics[scale=0.39]{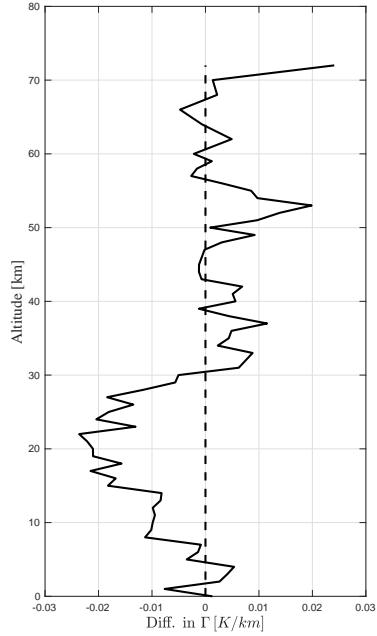}
	\vspace{-0.75cm}
	\caption{$\Gamma^{\text{Seiff1980}} - \Gamma^{\text{GERG2008}}$: Difference in adiabatic lapse rates computed for the VIRA model using the GERG-2008 model and that calculated in \citep{Seiff1985}}
	\label{fig:dev_adiabatic_lapse_rate_Seiff1985}
\end{figure}

An atmospheric model of Venus was created in \citep{Seiff1985} using the measurements obtained from the four Pioneer Venus probes. Details of the profiles measured by these probes can be found in \citep{Seiff1980}. The vertical profile of the adiabatic lapse rate for this atmospheric model was computed with the different mixture models which we have discussed using Eqs. \ref{eq:adiabatic_lapse_rate_HEOS} and \ref{eq:adiabatic_lapse_rate_IMRG}. The results for the GERG-2008 mixture model which was shown to be the most accurate mixture model against experimental data in Section \ref{sec:Verification-against-Experiments} are shown in Figure \ref{fig:lapse_rate_Seiff1985_Linkin1987}. The adiabatic lapse rate decreases with decreasing pressure and temperature from the surface by almost $1.5$ K/km between surface and $50$ km and increases by the same amount in the next $20$ km between $50$-$70$ km. Figure \ref{fig:dev_adiabatic_lapse_rate_Seiff1985} shows the difference between \cite{Seiff1985} adiabatic lapse rate computed for ideal mixture of $CO_2 - N_2$ (i.e. ignoring the real gas $CO_2-N_2$ interactions) and adiabatic lapse rate computed from the GERG-2008 model. The differences in the calculations are as high as $0.02$ K/km around $20$ km. This is high enough to characterize layers in the atmosphere close to neutrally stable, which were thought to be initially stable as now unstable. This shows the importance of taking non-ideal interactions in the real gas mixture into account and using more recent experimental data of $CO_2$ and $N_2$ represented by their equations of state.

\begin{figure}[H]
	\centering
	\centering
	\includegraphics[scale=0.4]{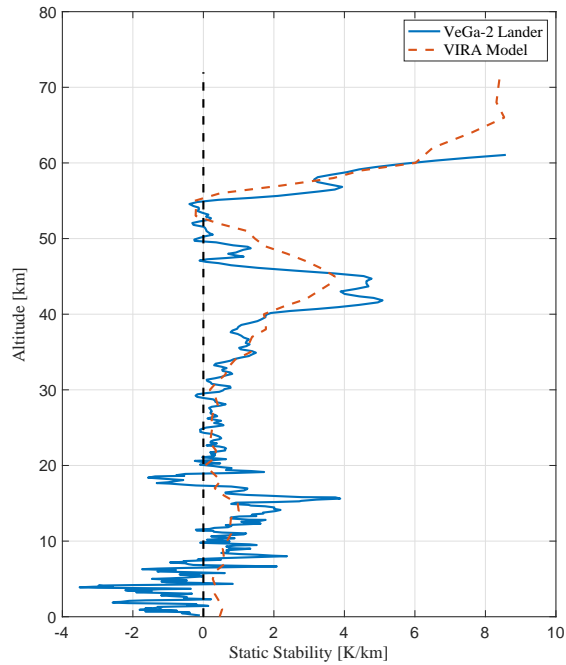}
	\caption{Comparison of profiles of static stability with altitude in the Venus atmosphere computed using the GERG-2008 mixture model considering the profiles measured by the VeGa-2 Lander \citep{Linkin1987} and the VIRA model \citep{Seiff1985} constructed from the four Pioneer Venus probes' data \citep{Seiff1980}}
	\label{fig:profiles_static_stability_Seiff1985_Linkin1987}
\end{figure}

VeGa 2 temperature profile \citep{Linkin1987} is the only one that provides measurements below $12$ km and the adiabatic lapse rates corresponding to this profile computed from the GERG-2008 model is shown in Figure \ref{fig:lapse_rate_Seiff1985_Linkin1987}. The corresponding static stability profiles are shown in Figure \ref{fig:profiles_static_stability_Seiff1985_Linkin1987} which was calculated using
\begin{equation}
\Delta\Gamma=\left(\frac{dT}{dz}\right)_{meas}-\left(\frac{dT}{dz}\right)_{ad} = \left(\frac{dT}{dz}\right)_{meas} + \Gamma
\end{equation}
where $\left(\frac{dT}{dz}\right)_{meas}$ is the gradient of the measured temperature with respect to altitude. This was computed using a second order centered scheme from the available temperature measurements. The nonlinear Savitzky-Golay filter was applied to the static stability profile computed for \citep{Linkin1987} with a span of $11$ points to remove spurious oscillations. Two superadiabatic layers are seen - one near the surface at about $4$ km and another at about $17$ km.  A layer of near neutral stability or even slightly unstable layer is also seen in the VeGa 2 profile between $50$-$54$ km. In the static stability profile for \citet{Seiff1985}, the atmosphere is stable near the surface. This difference between the static stability plots in Figure \ref{fig:profiles_static_stability_Seiff1985_Linkin1987} can be explained by looking at the difference in adiabatic lapse rates obtained in Figure \ref{fig:lapse_rate_Seiff1985_Linkin1987} near the surface for altitudes of $0-15$km.

\begin{figure}[H]
	\centering
	\centering
	\includegraphics[scale=0.4]{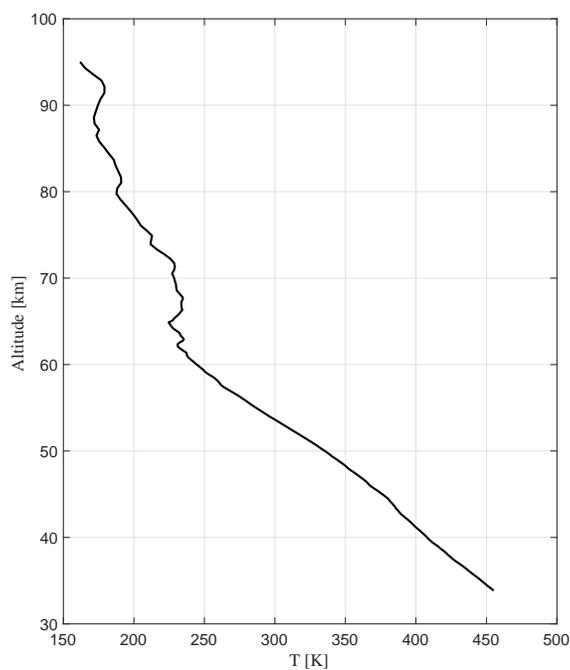}
	\caption{Profile of temperature with altitude of orbit 3212 of the Magellan spacecraft \citep{steffes1994radio,jenkins1994radio}}
	\label{fig:profile_Magellan1994}
\end{figure}

\begin{figure}[H]
	\centering
	\begin{subfigure}[t]{0.5\textwidth}
		\includegraphics[scale=0.4]{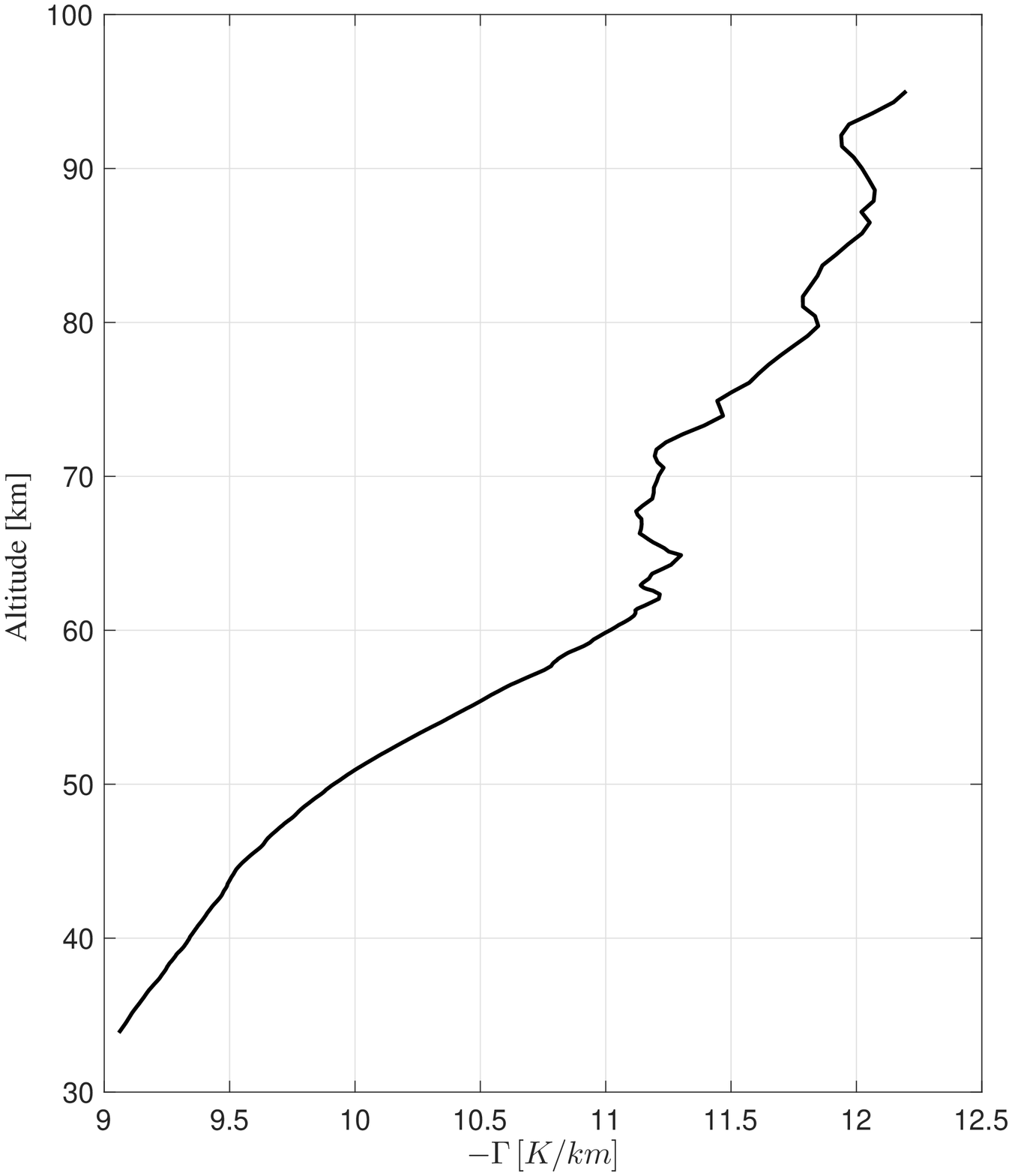}
		\caption{Profile of adiabatic lapse rate with altitude}
		\label{fig:adiabatic_lapse_rate_Magellan1994}
	\end{subfigure}%
	~
	\begin{subfigure}[t]{0.5\textwidth}
		\includegraphics[scale=0.4]{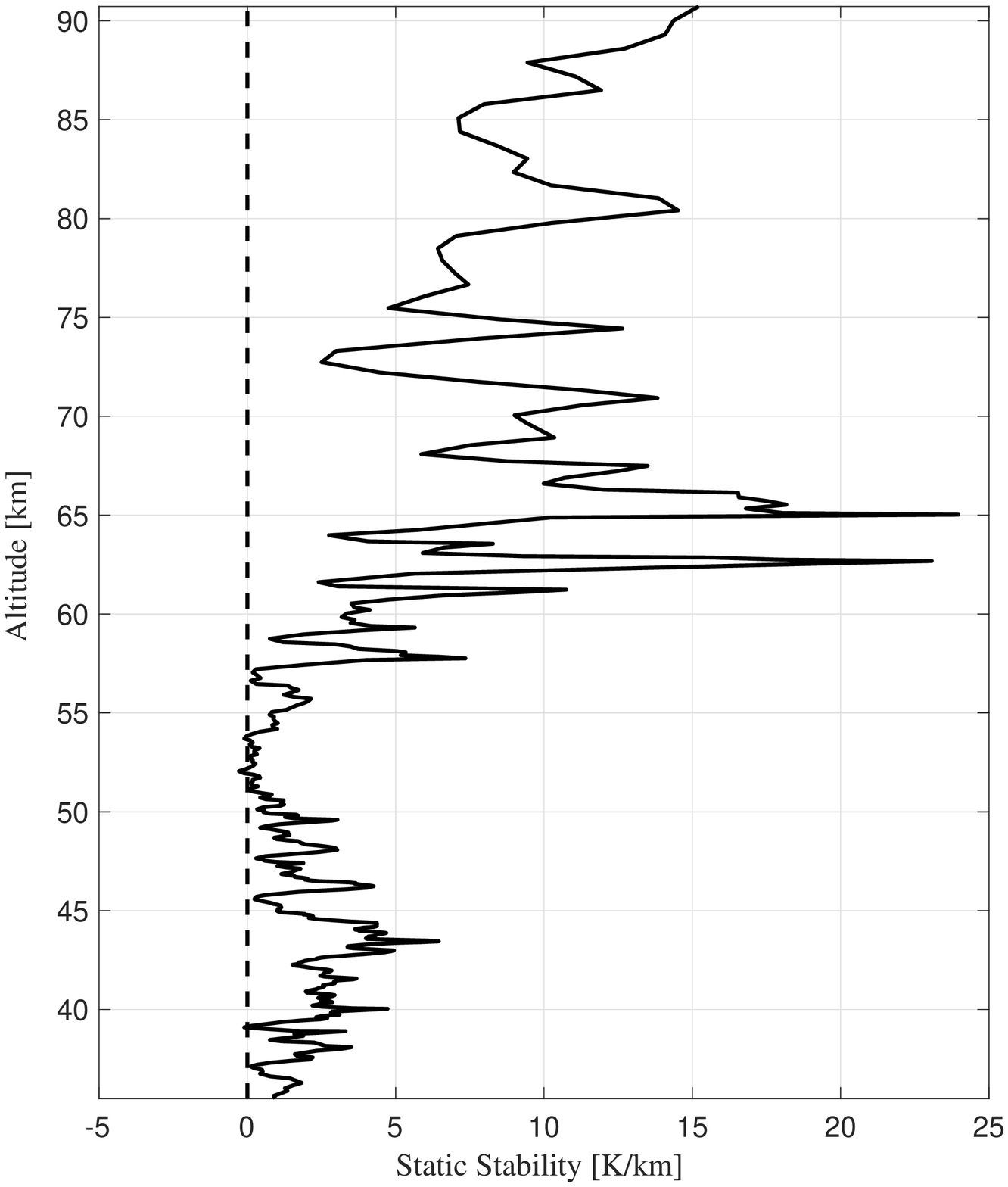}
		\caption{Profile of static stability with altitude}
		\label{fig:static_stability_Magellan1994}
	\end{subfigure}
	\caption{Profiles of adiabatic lapse rate and static stability with altitude in the Venus atmosphere considering the profile of orbit 3212 of the Magellan spacecraft \citep{steffes1994radio,jenkins1994radio} calculated using the GERG-2008 mixture model}
	\label{fig:Results_Magellan1994}
\end{figure}

For obtaining the adiabatic lapse rate and static stability for the higher altitudes in the Venus atmosphere, we use the $pT$ profiles (Figure \ref{fig:profile_Magellan1994}) obtained from radio occultation studies with the Magellan spacecraft \citep{steffes1994radio,jenkins1994radio}. The temperature profile used for the calculation of adiabatic lapse rate and static stability are from orbit 3212 of the spacecraft and is shown in Figure \ref{fig:profile_Magellan1994}. The results obtained for adiabatic lapse rate is shown in Figure \ref{fig:adiabatic_lapse_rate_Magellan1994} and for static stability is shown in Figure \ref{fig:static_stability_Magellan1994}. The vertical profile of static stability obtained using the GERG-2008 mixture model is similar to the one obtained in \citep{hinson1995magellan}. Differences are due to the fact that \citep{hinson1995magellan} used values of $\Gamma$ from \citep{Seiff1980}.

\begin{figure}[H]
	\centering
	\begin{subfigure}[t]{0.5\textwidth}
		\centering
		\includegraphics[scale=0.4]{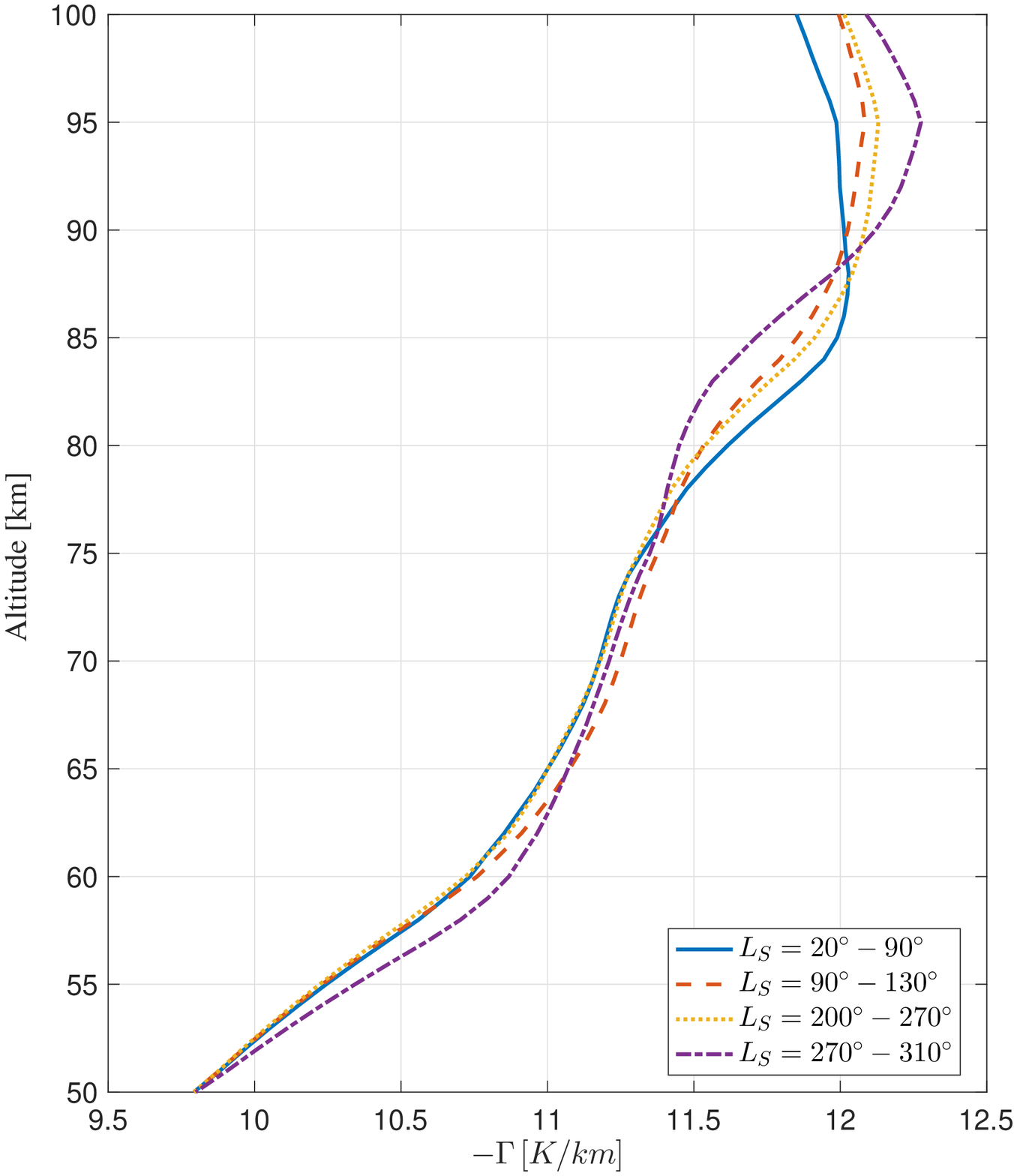}
		\caption{Profile of adiabatic lapse rate with altitude}
		\label{fig:adiabatic_lapse_rate_Zasova2006}
	\end{subfigure}%
	~
	\begin{subfigure}[t]{0.5\textwidth}
		\centering
		\includegraphics[scale=0.4]{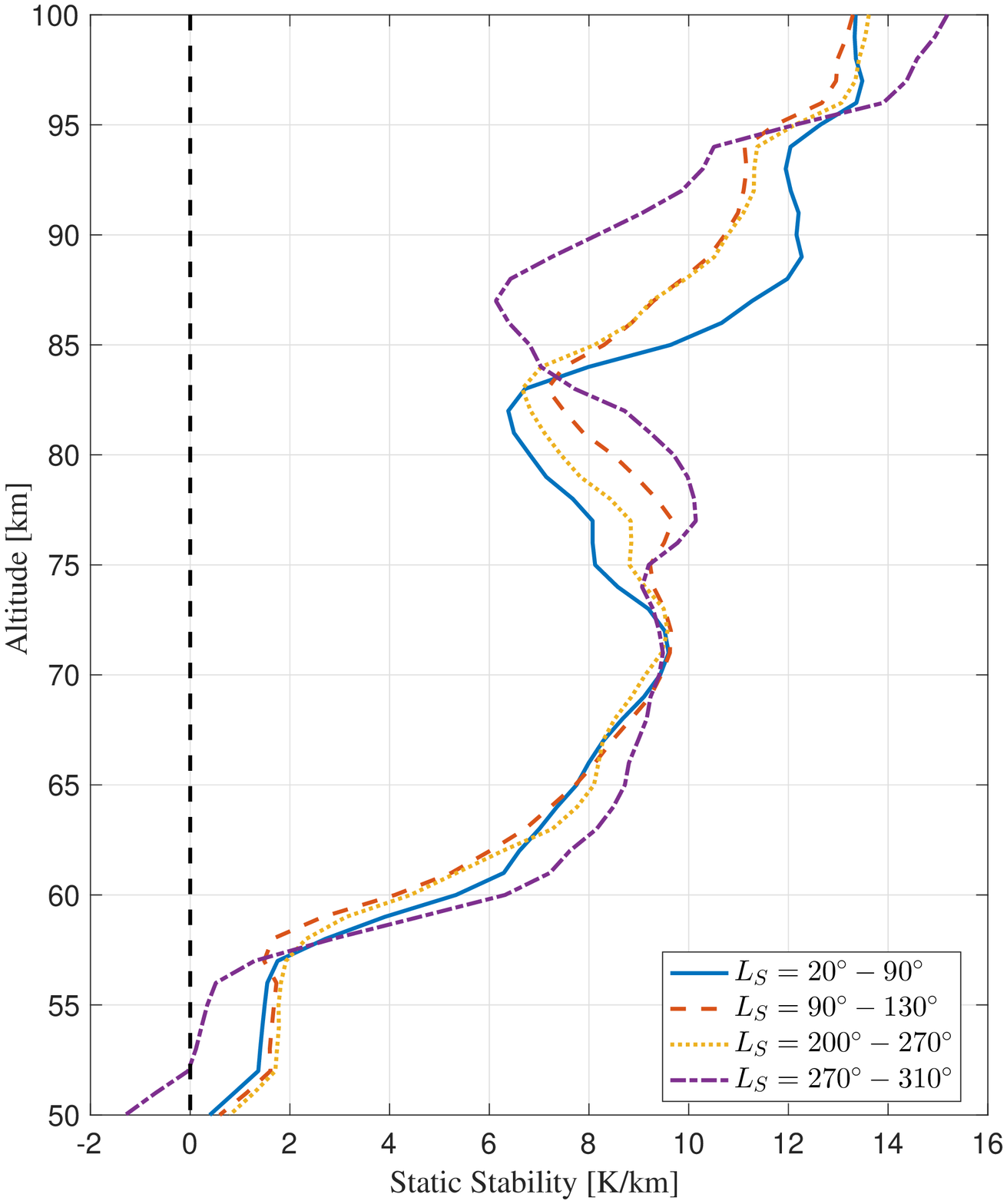}
		\caption{Profile of static stability with altitude}
		\label{fig:static_stability_Zasova2006}
	\end{subfigure}
	\caption{Profiles of adiabatic lapse rate and static stability with altitude as a function of solar longitude in the upper atmosphere of Venus atmosphere calculated using the GERG-2008 mixture model, considering the Venera 15 Fourier Spectrometer data \citep{Zasova2006}}
	\label{fig:Results_Zasova2006}
\end{figure}

From the infrared spectrometry data onboard Venera-15 \citep{Zasova2006}, it was observed that there are spatial and temporal variations in the upper atmosphere. To fully understand the convective stability in the Venus atmosphere, we take these into consideration when calculating adiabatic lapse rate and static stability. Figure \ref{fig:Results_Zasova2006} shows the profiles of adiabatic lapse rate and static stability for latitudes $\phi < 35^{\circ}$ and for various solar longitudes. Not only are there clear variations in the magnitude of static stability from $75 - 100$ km, we also observe that the atmosphere is unstable from $50-52$ km for solar longitude $L_S = 270^\circ - 310^\circ$ but stable otherwise. This indicates the importance of considering the variation in adiabatic lapse rate with both altitude and latitude.

\section{Conclusion and Future Work\label{sec:Conclusion-and-Future}}
We have calculated more accurate values of the adiabatic lapse rate for a mixture of $96.5\%$ carbon dioxide and $3.5\%$ nitrogen using the GERG-2008 mixture model for the temperature and pressure conditions found in the Venus atmosphere. We were able to account for the difference in adiabatic lapse rate values due to non-ideal interactions between $CO_2$ and $N_2$. Near the altitudes of $20\text{km}$, the magnitude of our value is about $0.02 \text{K/km}$ lower than the approximate value calculated by \cite{Seiff1980}. We showed the importance of considering spatial variations in adiabatic lapse rate with latitude and altitude as well as temporal variations. These calculations can also be performed considering the Venus atmosphere composition to vary with altitude to reflect the measured differences in the composition. It was shown in \citep{oyama1980pioneer} that the abundance of nitrogen in the atmosphere can be as high as $4.6 \, \text{v}\%$ at $51.6\text{km}$ and more recent studies \citep{peplowski2016nitrogen} have reported higher values of $5.38 \, \text{v}\%$ at $60-70\text{km}$. Further, considering the gradient in molecular weight with altitude will alter all available profiles of $T(p)$ for occultation and entry probe measurements, and $T(z)$ for non occultation results. Moreover, this approach can be applied to other planets or moons such as Saturn's largest moon Titan which has an atmosphere composed of mainly nitrogen and methane.

\section*{Acknowledgments}
Arkopal Dutt acknowledges support from the Indo US Science and Technology Foundation for the S.N. Bose Scholarship at University of Wisconsin-Madison. Funding from NASA Grant NNX09AE85G for completion of this work is acknowledged.

\appendix

\section{Derivation of Real Gas Adiabatic Lapse Rate}\label{Deriv_Gamma}
Here, we outline a derivation of the real gas adiabatic lapse rate along the lines of \citet{Staley1970} and show it is applicable to any planetary atmosphere with altitude varying real gas mixture composition. The system in consideration is a parcel of air composed of $m$ real gas components at altitude $z$. There are $n_{i}$ moles of component gas $i$. This parcel of air is engaged in the adiabatic process of rising in the atmosphere. The first law of thermodynamics gives us the following relationship between the internal energy $U$, heat $Q$ and work done $W$ on the system
\begin{equation}
dU=\delta Q-\delta W+\sum_{i}\mu_{i}dn_{i}
\end{equation}
where $\mu_{i}$ is the chemical potential of the $i$th component gas of the system and $dn_{i}$ is the change in number of moles of that particular component. Considering that the parcel of air has a constant composition while rising,
\begin{equation}
dU=\delta Q-\delta W\label{eq:first_law-1}
\end{equation}
Assuming specific internal energy $u$ as a function of temperature $T$, specific volume $v$ and composition we have
\begin{equation}
du=\left(\frac{\partial u}{\partial T}\right)_{v,\sum n_{i}}dT+\left(\frac{\partial u}{\partial v}\right)_{T,\sum n_{i}}dv+\sum_{j}\left(\frac{\partial u}{\partial n_{j}}\right)_{T,v,\sum_{i\neq j}n_{i}}dn_{j}
\end{equation}
where the subscript $\sum n_{i}$ denotes that the mole numbers of all the component gases is held constant for the corresponding partial derivative and $\sum_{j\neq i}n_{j}$ denotes that the mole numbers of all component gases but $j$th component is held constant. Noting that the composition of the air parcel does not change while rising, the above equation is simplified to
\begin{equation}
du=c_{v}dT+\left(\frac{\partial u}{\partial v}\right)_{T}dv
\end{equation}
where $c_{v}$ is the isochoric specific heat capacity. Using the Maxwell's relation of 
\begin{equation}
\left(\frac{\partial u}{\partial v}\right)_{T}=T\left(\frac{\partial p}{\partial T}\right)_{v}-p
\end{equation}
where $p$ is pressure in the parcel, we obtain from substituting in Eq.\ref{eq:first_law-1}:
\begin{equation}
dq=c_{v}dT+T\left(\frac{\partial p}{\partial T}\right)_{v}dv\label{eq:mod_firstlaw-1}
\end{equation}
If we were to introduce an equation of state explicit in pressure $p=p(v,T,n_{i})$ , we would then have
\begin{equation}
dp=\left(\frac{\partial p}{\partial T}\right)_{v,\sum n_{i}}dT+\left(\frac{\partial p}{\partial v}\right)_{T,\sum n_{i}}dv+\sum_{j}\left(\frac{\partial p}{\partial n_{j}}\right)_{T,v,\sum_{i\neq j}n_{i}}dn_{j}
\end{equation}
which reduces to 
\begin{equation}
dp=\left(\frac{\partial p}{\partial T}\right)_{v}dT+\left(\frac{\partial p}{\partial v}\right)_{T}dv
\end{equation}
as the parcel maintains constant composition while rising. Writing it explicitly in terms of volume differential, we have
\begin{equation}
dv=\frac{dp-\left(\frac{\partial p}{\partial T}\right)_{v}dT}{\left(\frac{\partial p}{\partial v}\right)_{T}}
\end{equation}
Substituting above in Eq.\ref{eq:mod_firstlaw-1}, we get
\begin{eqnarray}
dq & = & \left[c_{v}-\frac{T\left(\frac{\partial p}{\partial T}\right)_{v}^{2}}{\left(\frac{\partial p}{\partial v}\right)_{T}}\right]dT+T\frac{\left(\frac{\partial p}{\partial T}\right)_{v}}{\left(\frac{\partial p}{\partial v}\right)_{T}}dp\\
dq & = & c_{p}dT+T\frac{\left(\frac{\partial p}{\partial T}\right)_{v}}{\left(\frac{\partial p}{\partial v}\right)_{T}}dp
\end{eqnarray}
where $c_{p}$ is the isobaric specific heat capacity of the multi-component real gas mixture that makes up the parcel of air. Assuming adiabatic condition and using the hydrostatic equation $dp=-\rho gdz$ where
$\rho$ ($=1/v$) is the density of the air parcel, the following expression for adiabatic lapse rate $\Gamma$ is obtained.
\begin{equation}
\Gamma=-\frac{dT}{dz}=-T\rho\frac{\left(\frac{\partial p}{\partial T}\right)_{v}}{\left(\frac{\partial p}{\partial v}\right)_{T}}\left(\frac{g}{c_{p}}\right)=\frac{T}{\rho}\frac{\left(\frac{\partial p}{\partial T}\right)_{\rho}}{\left(\frac{\partial p}{\partial\rho}\right)_{T}}\left(\frac{g}{c_{p}}\right)
\end{equation}

\section{Equation of State for \texorpdfstring{$N_2$}{N2} \label{appendix_eos_N2}}

\begin{table}[ht]
	\begin{centering}
	\begin{tabular}{ccccc||ccccc}
		\hline 
		$k$ & $N_{k}$ & $i_{k}$ & $j_{k}$ & $l_{k}$ & $k$ & $N_{k}$ & $i_{k}$ & $j_{k}$ & $l_{k}$\tabularnewline
		\hline 
		1 & 0.924803575275  & 1 & 0.250  & 0 & 19 & -0.043576233605  & 1 & 4.000  & 2\tabularnewline
		2 & -0.492448489428  & 1 & 0.875  & 0 & 20 & -0.072317488932  & 2 & 6.000  & 2\tabularnewline
		3 & 0.661883336938  & 2 & 0.500  & 0 & 21 & 0.038964431527  & 3 & 6.000  & 2\tabularnewline
		4 & -1.929026492010  & 2 & 0.875  & 0 & 22 & -0.021220136391  & 4 & 3.000  & 2\tabularnewline
		5 & -0.062246930963  & 3 & 0.375  & 0 & 23 & 0.004088229815  & 5 & 3.000  & 2\tabularnewline
		6 & 0.349943957581  & 3 & 0.750  & 0 & 24 & -0.000055199002  & 8 & 6.000  & 2\tabularnewline
		7 & 0.564857472498  & 1 & 0.500  & 1 & 25 & -0.046201671648  & 4 & 16.000  & 3\tabularnewline
		8 & -1.617200059870  & 1 & 0.750  & 1 & 26 & -0.003003117160  & 5 & 11.000  & 3\tabularnewline
		9 & -0.481395031883  & 1 & 2.000  & 1 & 27 & 0.036882589121  & 5 & 15.000  & 3\tabularnewline
		10 & 0.421150636384  & 3 & 1.250  & 1 & 28 & -0.002558568462  & 8 & 12.000  & 3\tabularnewline
		11 & -0.016196223083  & 3 & 3.500  & 1 & 29 & 0.008969152646  & 3 & 12.000  & 4\tabularnewline
		12 & 0.172100994165  & 4 & 1.000  & 1 & 10 & -0.004415133704  & 5 & 7.000  & 4\tabularnewline
		13 & 0.007354489249  & 6 & 0.500  & 1 & 31 & 0.001337229249  & 6 & 4.000  & 4\tabularnewline
		14 & 0.016807730548  & 6 & 3.000  & 1 & 32 & 0.000264832492  & 9 & 16.000  & 4\tabularnewline
		15 & -0.001076266642  & 7 & 0.000  & 1 & 33 & 19.668819401500  & 1 & 0.000  & 2\tabularnewline
		16 & -0.013731808851  & 7 & 2.750  & 1 & 34 & -20.911560073000  & 1 & 1.000  & 2\tabularnewline
		17 & 0.000635466900  & 8 & 0.750  & 1 & 35 & 0.016778830699  & 3 & 2.000  & 2\tabularnewline
		18 & 0.003044322794  & 8 & 2.500  & 1 & 36 & 2627.675662740000  & 2 & 3.000  & 2\tabularnewline
		\hline 
	\end{tabular}\\
	~
	\par\end{centering}
	\vspace{1em}
	~
	\begin{centering}
	\begin{tabular}{cccc}
		\hline 
		$k$ & $\phi_{k}$ & $\beta_{k}$ & $\gamma_{k}$\tabularnewline
		\hline 
		33 & 20 & 325 & 1.16\tabularnewline
		34 & 20 & 325 & 1.16\tabularnewline
		35 & 15 & 300 & 1.13\tabularnewline
		36 & 25 & 275 & 1.25\tabularnewline
		\hline 
	\end{tabular}
	\par\end{centering}
\protect\caption{Parameters as in Eq.\ref{eq:alpha_r_N2}}
\end{table}

Derivatives required for the evaluation of thermodynamic properties are given by
\begin{dmath}
	\left(\frac{\partial^{2}\alpha_{N_{2}}^{0}}{\partial\tau^{2}}\right)_{\delta}=-a_{1}\tau^{-2}+2a_{4}\tau^{-3}+6a_{5}\tau^{-4}+12a_{6}\tau^{-5}-a_{7}a_{8}\frac{\exp(a_{8}\tau)}{[\exp(a_{8}\tau)-1]^{2}}
\end{dmath}
\begin{dmath}
	\left(\frac{\partial\alpha_{N_{2}}^{r}}{\partial\delta}\right)_{\tau} = \sum_{k=1}^{6}i_{k}N_{k}\delta^{i_{k}-1}\tau^{j_{k}}+\sum_{k=7}^{32}N_{k}\delta^{i_{k}-1}\tau^{j_{k}}\exp(-\delta^{l_{k}})(i_{k}-l_{k}\delta^{l_{k}}) +\sum_{k=33}^{36}N_{k}\delta^{i_{k}-1}\tau^{j_{k}}\exp(-\phi_{k}(\delta-1)^{2}-\beta_{k}(\tau-\gamma_{k})^{2})[i_{k}-2\delta\phi_{k}(\delta-1)]
\end{dmath}
\begin{dmath}
	\left(\frac{\partial^{2}\alpha_{N_{2}}^{r}}{\partial\delta^{2}}\right)_{\tau} = \sum_{k=1}^{6}i_{k}(i_{k}-1)N_{k}\delta^{i_{k}-2}\tau^{j_{k}} +\sum_{k=7}^{32}N_{k}\delta^{i_{k}-2}\tau^{j_{k}}\exp(-\delta^{l_{k}})[(i_{k}-l_{k}\delta^{l_{k}})(i_{k}-1-l_{k}\delta^{l_{k}})-l_{k}^{2}\delta^{l_{k}}] +\sum_{k=33}^{36}N_{k}\delta^{i_{k}-2}\tau^{j_{k}}\exp(-\phi_{k}(\delta-1)^{2}-\beta_{k}(\tau-\gamma_{k})^{2})\{[i_{k}-2\phi_{k}\delta(\delta-1)]^{2}-i_{k}-2\delta^{2}\phi_{k}\}
\end{dmath}
\begin{dmath}
	\left(\frac{\partial^{2}\alpha_{N_{2}}^{r}}{\partial\delta\partial\tau}\right) = \sum_{k=1}^{6}i_{k}j_{k}N_{k}\delta^{i_{k}-1}\tau^{j_{k}-1}+\sum_{k=7}^{32}j_{k}N_{k}\delta^{i_{k}-1}\tau^{j_{k}-1}\exp(-\delta^{l_{k}})(i_{k}-l_{k}\delta^{l_{k}}) +\sum_{k=33}^{36}N_{k}\delta^{i_{k}-1}\tau^{j_{k}-1}\exp(-\phi_{k}(\delta-1)^{2}-\beta_{k}(\tau-\gamma_{k})^{2})[i_{k}-2\delta\phi_{k}(\delta-1)][j_{k}-2\tau\beta_{k}(\tau-\gamma_{k})]
\end{dmath}
\begin{dmath}
	\left(\frac{\partial^{2}\alpha_{N_{2}}^{r}}{\partial\tau^{2}}\right)_{\delta} = \sum_{k=1}^{6}j_{k}(j_{k}-1)N_{k}\delta^{i_{k}}\tau^{j_{k}-2}+\sum_{k=7}^{32}j_{k}(j_{k}-1)N_{k}\delta^{i_{k}}\tau^{j_{k}-2}\exp(-\delta^{l_{k}}) +\sum_{k=33}^{36}N_{k}\delta^{i_{k}}\tau^{j_{k}-2}\exp(-\phi_{k}(\delta-1)^{2}-\beta_{k}(\tau-\gamma_{k})^{2})\{[j_{k}-2\beta_{k}\tau(\tau-\gamma_{k})]^{2}-j_{k}-2\tau^{2}\beta_{k}\}
\end{dmath}

\section{Equation of State for \texorpdfstring{$CO_2$}{CO2} \label{appendix_eos_CO2}}

\begin{table}[ht]
\begin{centering}
	\begin{tabular}{crrrr||rrrrr}
	\hline 
	$i$ & $n_{i}$ & $d_{i}$ & $t_{i}$ & $c_{i}$ & $i$ & $n_{i}$ & $d_{i}$ & $t_{i}$ & $c_{i}$\tabularnewline
	\hline 
	1 & 0.38856823203161 & 1 & 0.000 &  & 18 & -0.01677587970043 & 1 & 6.000 & 2\tabularnewline
	2 & 2.93854759427400 & 1 & 0.750 &  & 19 & -0.11960736637987 & 4 & 3.000 & 2\tabularnewline
	3 & -5.58671885349340 & 1 & 1.000 &  & 20 & -0.04561936250878 & 4 & 6.000 & 2\tabularnewline
	4 & -0.76753199592477 & 1 & 2.000 &  & 21 & 0.03561278927035 & 4 & 8.000 & 2\tabularnewline
	5 & 0.31729005580416 & 2 & 0.750 &  & 22 & -0.00744277271321 & 7 & 6.000 & 2\tabularnewline
	6 & 0.54803315897767 & 2 & 2.000 &  & 23 & -0.00173957049024 & 8 & 0.000 & 2\tabularnewline
	7 & 0.12279411220335 & 3 & 0.750 &  & 24 & -0.02181012128953 & 2 & 7.000 & 3\tabularnewline
	8 & 2.16589615432200 & 1 & 1.500 & 1 & 25 & 0.02433216655924 & 3 & 12.000 & 3\tabularnewline
	9 & 1.58417351097240 & 2 & 1.500 & 1 & 26 & -0.03744013342346 & 3 & 16.000 & 3\tabularnewline
	10 & -0.23132705405503 & 4 & 2.500 & 1 & 27 & 0.14338715756878 & 5 & 22.000 & 4\tabularnewline
	11 & 0.05811691643144 & 5 & 0.000 & 1 & 28 & -0.13491969083286 & 5 & 24.000 & 4\tabularnewline
	12 & -0.55369137205382 & 5 & 1.500 & 1 & 29 & -0.02315122505348 & 6 & 16.000 & 4\tabularnewline
	13 & 0.48946615909422 & 5 & 2.000 & 1 & 30 & 0.01236312549290 & 7 & 24.000 & 4\tabularnewline
	14 & -0.02427573984350 & 6 & 0.000 & 1 & 31 & 0.00210583219729 & 8 & 8.000 & 4\tabularnewline
	15 & 0.06249479050168 & 6 & 1.000 & 1 & 32 & -0.00033958519026 & 10 & 2.000 & 4\tabularnewline
	16 & -0.12175860225246 & 6 & 2.000 & 1 & 33 & 0.00559936517716 & 4 & 28.000 & 5\tabularnewline
	17 & -0.37055685270086 & 1 & 3.000 & 2 & 34 & -0.00030335118056 & 8 & 14.000 & 6\tabularnewline
	\hline 
	\end{tabular}
\par\end{centering}
\vspace{1em}
\begin{centering}
	\begin{centering}
	\begin{tabular}{crrrrrrr}
	\hline 
	$i$ & $n_{i}$ & $d_{i}$ & $t_{i}$ & $\alpha_{i}$ & $\beta_{i}$ & $\gamma_{i}$ & $\epsilon_{i}$\tabularnewline
	\hline 
	35 & -213.65488688320000 & 2 & 1.000 & 25 & 325 & 1.16 & 1\tabularnewline
	36 & 26641.56914927200000 & 2 & 0.000 & 25 & 300 & 1.19 & 1\tabularnewline
	37 & -24027.21220455700000 & 2 & 1.000 & 25 & 300 & 1.19 & 1\tabularnewline
	38 & -283.41603423999000 & 3 & 3.000 & 15 & 275 & 1.25 & 1\tabularnewline
	39 & 212.47284400179000 & 3 & 3.000 & 20 & 275 & 1.22 & 1\tabularnewline
	\hline 
	\end{tabular}
	\par\end{centering}
\par\end{centering}
\vspace{1em}
\begin{centering}
	\begin{centering}
	\begin{tabular}{crrrrrrrr}
	\hline 
	$i$ & $n_{i}$ & $a_{i}$ & $b_{i}$ & $\beta_{i}$ & $A_{i}$ & $B_{i}$ & $C_{i}$ & $D_{i}$\tabularnewline
	\hline 
	40 & -0.66642276540751 & 3.5 & 0.875 & 0.3 & 0.7 & 0.3 & 10 & 275\tabularnewline
	41 & 0.72608632349897 & 3.5 & 0.925 & 0.3 & 0.7 & 0.3 & 10 & 275\tabularnewline
	42 & 0.05506866861284 & 3 & 0.875 & 0.3 & 0.7 & 1 & 12.5 & 275\tabularnewline
	\hline 
	\end{tabular}
	\par\end{centering}
\par\end{centering}
\protect\caption{Parameters as in Eq.\ref{eq:alpha_r_CO2}}
\end{table}

Derivatives as required are given by
\begin{dmath}
	\left(\frac{\partial\alpha_{CO_{2}}^{r}}{\partial\delta}\right)_{\tau} = \sum_{i=1}^{7}n_{i}d_{i}\delta^{d_{i}-1}\tau^{t_{i}}+\sum_{i=8}^{34}n_{i}\exp(-\delta^{c_{i}})[\delta^{d_{i}-1}\tau^{t_{i}}(d_{i}-c_{i}\delta^{c_{i}})] +\sum_{i=35}^{39}n_{i}\delta^{d_{i}}\tau^{t_{i}}\exp(-\alpha_{i}(\delta-\epsilon_{i})^{2}-\beta_{i}(\tau-\gamma_{i})^{2})\left[\frac{d_{i}}{\delta}-2\alpha_{i}(\delta-\epsilon_{i})\right] +\sum_{i=40}^{42}n_{i}\left[\Delta^{b_{i}}\left(\Psi+\delta\frac{\partial\Psi}{\partial\delta}\right)+\frac{\partial\Delta^{b_{i}}}{\partial\delta}\delta\Psi\right]
\end{dmath}
\begin{dmath}
	\left(\frac{\partial^{2}\alpha_{CO_{2}}^{r}}{\partial\delta^{2}}\right)_{\tau} = \sum_{i=1}^{7}n_{i}d_{i}(d_{i}-1)\delta^{d_{i}-2}\tau^{t_{i}}+\sum_{i=8}^{34}n_{i}\exp(-\delta^{c_{i}})[\delta^{d_{i}-2}\tau^{t_{i}}((d_{i}-c_{i}\delta^{c_{i}})(d_{i}-1-c_{i}\delta^{c_{i}})-c_{i}^{2}\delta^{c_{i}}] +\sum_{i=35}^{39}n_{i}\tau^{t_{i}}\exp(-\alpha_{i}(\delta-\epsilon_{i})^{2}-\beta_{i}(\tau-\gamma_{i})^{2})\times [-2\alpha_{i}\delta^{d_{i}}+4\alpha_{i}^{2}\delta^{d_{i}-1}(\delta-\epsilon_{i})^{2}-4d_{i}\alpha_{i}\delta^{d_{i}-1}(\delta-\epsilon_{i})+d_{i}(d_{i}-1)\delta^{d_{i}-2}] +\sum_{i=40}^{42}n_{i}\left[\Delta^{b_{i}}\left(2\frac{\partial\Psi}{\partial\delta}+\delta\frac{\partial^{2}\Psi}{\partial\delta^{2}}\right)+2\frac{\partial\Delta^{b_{i}}}{\partial\delta}\left(\Psi+\delta\frac{\partial\Psi}{\partial\delta}\right)+\frac{\partial^{2}\Delta^{b_{i}}}{\partial\delta^{2}}\delta\Psi\right]
\end{dmath}
\begin{dmath}
	\left(\frac{\partial^{2}\alpha_{CO_{2}}^{r}}{\partial\delta\partial\tau}\right) = \sum_{i=1}^{7}n_{i}d_{i}t_{i}\delta^{d_{i}-1}\tau^{t_{i}-1}+\sum_{i=8}^{34}n_{i}\exp(-\delta^{c_{i}})\delta^{d_{i}-1}t_{i}\tau^{t_{i}-1}(d_{i}-c_{i}\delta^{c_{i}}) +\sum_{i=35}^{39}n_{i}\delta^{d_{i}}\tau^{t_{i}}\exp(-\alpha_{i}(\delta-\epsilon_{i})^{2}-\beta_{i}(\tau-\gamma_{i})^{2})\left[\frac{d_{i}}{\delta}-2\alpha_{i}(\delta-\epsilon_{i})\right]\left[\frac{t_{i}}{\tau}-2\beta_{i}(\tau-\gamma_{i})\right] +\sum_{i=40}^{42}n_{i}\left[\Delta^{b_{i}}\left(\frac{\partial\Psi}{\partial\tau}+\delta\frac{\partial^{2}\Psi}{\partial\delta\partial\tau}\right)+\delta\frac{\partial\Delta^{b_{i}}}{\partial\delta}\frac{\partial\Psi}{\partial\tau}+\frac{\partial\Delta^{b_{i}}}{\partial\tau}\left(\Psi+\delta\frac{\partial\Psi}{\partial\delta}\right)+\frac{\partial^{2}\Delta^{b_{i}}}{\partial\delta\partial\tau}\delta\Psi\right]
\end{dmath}
\begin{dmath}
	\left(\frac{\partial^{2}\alpha_{CO_{2}}^{r}}{\partial\tau^{2}}\right)_{\delta} = \sum_{i=1}^{7}n_{i}t_{i}(t_{i}-1)\delta^{d_{i}}\tau^{t_{i}-1}+\sum_{i=8}^{34}n_{i}t_{i}(t_{i}-1)\delta^{d_{i}}\tau^{t_{i}-2}\exp(-\delta^{c_{i}}) +\sum_{i=35}^{39}n_{i}\delta^{d_{i}}\tau^{t_{i}}\exp(-\alpha_{i}(\delta-\epsilon_{i})^{2}-\beta_{i}(\tau-\gamma_{i})^{2})\left[\left(\frac{t_{i}}{\tau}-2\beta_{i}(\tau-\gamma_{i})\right)^{2}-\frac{t_{i}}{\tau^{2}}-2\beta_{i}\right] +\sum_{i=40}^{42}n_{i}\delta\left[\frac{\partial^{2}\Delta^{b_{i}}}{\partial\tau^{2}}\Psi+2\frac{\partial\Delta^{b_{i}}}{\partial\tau}\frac{\partial\Psi}{\partial\tau}+\Delta^{b_{i}}\frac{\partial^{2}\Psi}{\partial\tau^{2}}\right]
\end{dmath}
Derivatives of $\Delta$ and $\Delta^{b_{i}}$ are given by
\begin{equation}
\frac{\partial\Delta}{\partial\delta}=(\delta-1)\left\{ A_{i}\theta\frac{2}{\beta_{i}}[(\delta-1)^{2}]^{\frac{1}{2\beta_{i}}-1}+2B_{i}a_{i}[(\delta-1)^{2}]^{a_{i}-1}\right\} 
\end{equation}
\begin{dmath}
	\frac{\partial^{2}\Delta}{\partial\delta^{2}} = \frac{1}{\delta-1}\frac{\partial\Delta}{\partial\delta}+(\delta-1)^{2}\times \left\{ 4B_{i}a_{i}(a_{i}-1)[(\delta-1)^{2}]^{a_{i}-2}+2A_{i}^{2}\left(\frac{1}{\beta_{i}}\right)^{2}\{[(\delta-1)^{2}]^{\frac{1}{2\beta_{i}}-1}\}^{2}+A_{i}\theta\frac{4}{\beta_{i}}\left(\frac{1}{2\beta_{i}}-1\right)[(\delta-1)^{2}]^{\frac{1}{2\beta_{i}}-2}\right\} \\
\end{dmath}
\begin{eqnarray}
\frac{\partial\Delta^{b_{i}}}{\partial\delta} & = & b_{i}\Delta^{b_{i}-1}\frac{\partial\Delta}{\partial\delta} \\
\frac{\partial^{2}\Delta^{b_{i}}}{\partial\delta^{2}} & = & b_{i}\left[\Delta^{b_{i}-1}\frac{\partial^{2}\Delta}{\partial\delta^{2}}+(b_{i}-1)\Delta^{b_{i}-2}\left(\frac{\partial\Delta}{\partial\delta}\right)^{2}\right]\\
\frac{\partial\Delta^{b_{i}}}{\partial\tau} & = & -2\theta b_{i}\Delta^{b_{i}-1}\\
\frac{\partial^{2}\Delta^{b_{i}}}{\partial\tau^{2}} & = & 2b_{i}\Delta^{b_{i}-1}+4\theta^{2}b_{i}(b_{i}-1)\Delta^{b_{i}-2}\\
\frac{\partial^{2}\Delta^{b_{i}}}{\partial\delta\partial\tau} & = & -A_{i}b_{i}\frac{2}{\beta_{i}}\Delta^{b_{i}-1}(\delta-1)[(\delta-1)^{2}]^{\frac{1}{2\beta_{i}}-1}-2\theta b_{i}(b_{i}-1)\Delta^{b_{i}-2}\frac{\partial\Delta}{\partial\delta}
\end{eqnarray}
Derivatives of $\Psi$ are given by
\begin{eqnarray}
\frac{\partial\Psi}{\partial\delta} & = & -2C_{i}\Psi(\delta-1)\\
\frac{\partial^{2}\Psi}{\partial\delta^{2}} & = & 2C_{i}\Psi[2C_{i}(\delta-1)^{2}-1]\\
\frac{\partial\Psi}{\partial\tau} & = & -2D_{i}\Psi(\tau-1)\\
\frac{\partial^{2}\Psi}{\partial\tau^{2}} & = & 2D_{i}\Psi[2D_{i}(\tau-1)^{2}-1]\\
\frac{\partial^{2}\Psi}{\partial\delta\partial\tau} & = & 4C_{i}D_{i}\Psi(\delta-1)(\tau-1)
\end{eqnarray}


\bibliographystyle{elsarticle-harv}
\bibliography{paper}

\begin{thebibliography}{32}
\expandafter\ifx\csname natexlab\endcsname\relax\def\natexlab#1{#1}\fi
\expandafter\ifx\csname url\endcsname\relax
  \def\url#1{\texttt{#1}}\fi
\expandafter\ifx\csname urlprefix\endcsname\relax\def\urlprefix{URL }\fi

\bibitem[{Bishnoi et~al.(1972)Bishnoi, Hamaliuk, and Robinson}]{Bishnoi1972}
Bishnoi, P.~R., Hamaliuk, G.~P., Robinson, D.~B., October 1972. Experimental
  {H}eat {C}apacities of {N}itrogen-{C}arbon {D}ioxide {M}ixtures at {E}levated
  {P}ressures. The Canadian Journal of Chemical Engineering 50, 677--679.

\bibitem[{Brugge et~al.(1989)Brugge, Hwang, Rogers, Holste, Hall, Lemming,
  Esper, Marsh, and Gammon}]{Brugge1989}
Brugge, H., Hwang, C.-A., Rogers, W., Holste, J., Hall, K., Lemming, W., Esper,
  G., Marsh, K., Gammon, B., 1989. Experimental {C}ross {V}irial {C}oefficients
  for {B}inary {M}ixtures of {C}arbon {D}ioxide with {N}itrogen, {M}ethane and
  {E}thane at 300 and 320 k. Physica A 156, 382--416.

\bibitem[{Brugge et~al.(1997)Brugge, Holste, Hall, Gammon, and
  Marsh}]{Brugge1997}
Brugge, H.~B., Holste, J.~C., Hall, K.~R., Gammon, B.~E., Marsh, K.~N., 1997.
  Densities of {C}arbon {D}ioxide + {N}itrogen from 225 {K} to 450 {K} at
  {P}ressures up to 70 {MPa}. Journal of Chemical Engineering Data 42~(5),
  903--907.

\bibitem[{Ely et~al.(1989)Ely, Haynes, and Bain}]{Ely1989}
Ely, J., Haynes, W., Bain, B., 1989. Isochoric {(p,$V_{m}$,T)} measurements on
  ${CO}_{2}$ and on (0.982${CO}_2$ + 0.018${N}_2$) from 250 to 350 {K} at
  pressures to 35 {MPa}. Journal of Chemical Thermodynamics 21, 879--894.

\bibitem[{Gernert(2013)}]{gernert2013new}
Gernert, G.~J., 2013. A new helmholtz energy model for humid gases and ccs
  mixtures. Ph.D. thesis.

\bibitem[{Gernert et~al.(2014)Gernert, J{\"a}ger, and
  Span}]{gernert2014calculation}
Gernert, J., J{\"a}ger, A., Span, R., 2014. Calculation of phase equilibria for
  multi-component mixtures using highly accurate helmholtz energy equations of
  state. Fluid Phase Equilibria 375, 209--218.

\bibitem[{Hagermann et~al.(2007)Hagermann, Rosenberg, Towner, Garry, Svedhem,
  Leese, Hathi, Lorenz, and Zarnecki}]{hagermann2007speed}
Hagermann, A., Rosenberg, P., Towner, M., Garry, J., Svedhem, H., Leese, M.,
  Hathi, B., Lorenz, R., Zarnecki, J., 2007. Speed of sound measurements and
  the methane abundance in titan's atmosphere. Icarus 189~(2), 538--543.

\bibitem[{Hilsenrath et~al.(1955)Hilsenrath, Beckett, Benedict, and
  Lilla~Fano}]{Hilsenrath1955}
Hilsenrath, J., Beckett, C.~W., Benedict, W.~S., Lilla~Fano, Harold J.~Hoge, J.
  F. M. R. L. N. Y. S. T. H. W.~W., 1955. Tables of {T}hermal {P}roperties of
  {G}ases. National Bureau of Standards.

\bibitem[{Hinson and Jenkins(1995)}]{hinson1995magellan}
Hinson, D.~P., Jenkins, J.~M., 1995. Magellan radio occultation measurements of
  atmospheric waves on venus. Icarus 114~(2), 310--327.

\bibitem[{Jenkins et~al.(1994)Jenkins, Steffes, Hinson, Twicken, and
  Tyler}]{jenkins1994radio}
Jenkins, J.~M., Steffes, P.~G., Hinson, D.~P., Twicken, J.~D., Tyler, G.~L.,
  1994. Radio occultation studies of the venus atmosphere with the magellan
  spacecraft: 2. results from the october 1991 experiments. Icarus 110~(1),
  79--94.

\bibitem[{Klimeck(1996)}]{klimeck2000entwicklung}
Klimeck, R., May 1996. Entwicklung einer fundamentalgleichung f{\"u}r erdgase
  f{\"u}r das gas-und fl{\"u}ssigkeitsgebiet sowie das phasengleichgewicht.
  Ph.D. thesis, Ruhr-Universit{\"a}t Bochum.

\bibitem[{Kunz et~al.(2007)Kunz, Klimeck, Wagner, and Jaeschke}]{Kunz2004}
Kunz, O., Klimeck, R., Wagner, W., Jaeschke, M., 2007. The {GERG-2004}
  wide-range equation of state for natural gases and other mixtures. GERG {TM}
  15 6~(557).

\bibitem[{Kunz and Wagner(2012)}]{Kunz2012}
Kunz, O., Wagner, W., 2012. The {GERG-2008} wide-range equation of state for
  natural gases and other mixtures: an expansion of gerg-2004. Journal of
  {C}hemical \& {E}ngineering {D}ata 57~(11), 3032--3091.

\bibitem[{Lemmon(1996)}]{Lemmon1996}
Lemmon, E., May 1996. A {G}eneralized {M}odel for the {P}rediction of the
  {T}hermodynamic {P}roperties of {M}ixtures including {V}apor-{L}iquid
  {E}quilibrium. Ph.D. thesis, University of Idaho.

\bibitem[{Lemmon and Jacobsen(1999)}]{Lemmon1999}
Lemmon, E., Jacobsen, R., 1999. A {G}eneralized {M}odel for the {T}hermodynamic
  {P}roperties of {M}ixtures. International Journal of Thermophysics 20~(3),
  825--835.

\bibitem[{Linkin et~al.(1987)Linkin, Blamont, Devyatkin, Ignatova,
  Kerzhanovich, Lipatov, Malik, Stadnyk, Sanotskii, Stolyarchuk,
  et~al.}]{Linkin1987}
Linkin, V., Blamont, J., Devyatkin, S., Ignatova, S., Kerzhanovich, V.,
  Lipatov, A., Malik, K., Stadnyk, B., Sanotskii, Y.~V., Stolyarchuk, P.,
  et~al., 1987. Thermal structure of the atmosphere of venus from the results
  of measurements taken by landing vehicle vega-2. Kosnicheskie Issledovaniya
  25~(5), 659--672.

\bibitem[{Mantovani et~al.(2012)Mantovani, Chiesa, Valenti, Gatti, and
  Consonni}]{Mantovani2012}
Mantovani, M., Chiesa, P., Valenti, G., Gatti, M., Consonni, S., 2012.
  Supercritical pressure--density--temperature measurements on
  ${CO}_{2}$--$n_2$, ${CO}_{2}$--$o_2$ and ${CO}_{2}$--$ar$ binary mixtures.
  The Journal of Supercritical Fluids 61, 34--43.

\bibitem[{Moroz(1981)}]{moroz1981atmosphere}
Moroz, V., 1981. The atmosphere of venus. Space Science Reviews 29~(1), 3--127.

\bibitem[{Oyama et~al.(1980)Oyama, Carle, Woeller, Pollack, Reynolds, and
  Craig}]{oyama1980pioneer}
Oyama, V., Carle, G., Woeller, F., Pollack, J., Reynolds, R., Craig, R., 1980.
  Pioneer venus gas chromatography of the lower atmosphere of venus. Journal of
  Geophysical Research: Space Physics 85~(A13), 7891--7902.

\bibitem[{Peplowski and Lawrence(2016)}]{peplowski2016nitrogen}
Peplowski, P., Lawrence, D., 2016. Nitrogen content of venus' upper atmosphere
  from the messenger neutron spectrometer. In: Lunar and Planetary Science
  Conference. Vol.~47. p. 1177.

\bibitem[{Seiff et~al.(1980)Seiff, Kirk, Young, Blanchard, Findlay, Kelley, and
  Sommer}]{Seiff1980}
Seiff, A., Kirk, D.~B., Young, R.~E., Blanchard, R.~C., Findlay, J.~T., Kelley,
  G.~M., Sommer, S.~C., December 1980. Measurements of {T}hermal {S}tructure
  and {T}hermal {C}onstants in the {A}tmosphere of {V}enus and {R}elated
  {D}ynamical {O}bservations: {R}esults {F}rom the {F}our {P}ioneer {V}enus
  {P}robes. Journal of Geophysical Research 85~(A13), 7903--7933.

\bibitem[{Seiff et~al.(1985)Seiff, Schofield, Kliore, Taylor, Limaye,
  Revercomb, Sromovsky, Kerzhanovich, Moroz, and Marov}]{Seiff1985}
Seiff, A., Schofield, J., Kliore, A., Taylor, F., Limaye, S., Revercomb, H.,
  Sromovsky, L., Kerzhanovich, V., Moroz, V., Marov, M.~Y., 1985. Models of the
  structure of the atmosphere of venus from the surface to 100 kilometers
  altitude. Advances in Space Research 5~(11), 3--58.

\bibitem[{Sengers et~al.(2000)Sengers, Kayser, Peters, and
  White~Jr.}]{Sengers2000}
Sengers, J., Kayser, R., Peters, C., White~Jr., J. (Eds.), 2000. Equations of
  {S}tate for {F}luids and {F}luid {M}ixtures. Vol.~5 of Experimental
  Thermodynamics. Elsevier.

\bibitem[{Span et~al.(2000)Span, Lemmon, Jacobsen, Wagner, and
  Yokozeki}]{Span2000}
Span, R., Lemmon, E.~W., Jacobsen, R.~T., Wagner, W., Yokozeki, A., 2000. A
  {R}eference {E}quation of {S}tate for the {T}hermodynamic {P}roperties of
  {N}itrogen for {T}emperatures from 63.151 to 1000 {K} and {P}ressure to 2200
  {MPa}. Journal of Physical and Chemical Reference Data 29~(6), 1361--1433.

\bibitem[{Span and Wagner(1996)}]{Span1996}
Span, R., Wagner, W., 1996. A {N}ew {E}quation of {S}tate for {C}arbon
  {D}ioxide {C}overing the {F}luid {R}egion from the {T}riple-{P}oint
  {T}emperature to 1100 {K} at {P}ressures up to 800 {MPa}. Journal of Physical
  and Chemical Reference Data 25~(6).

\bibitem[{Staley(1970)}]{Staley1970}
Staley, D.~O., 1970. The {A}diabatic {L}apse {R}ate in the {V}enus
  {A}tmosphere. Journal of the Atmospheric Sciences 27, 219--223.

\bibitem[{Steffes et~al.(1994)Steffes, Jenkins, Austin, Asmar, Lyons, Seale,
  and Tyler}]{steffes1994radio}
Steffes, P.~G., Jenkins, J.~M., Austin, R.~S., Asmar, S.~W., Lyons, D.~T.,
  Seale, E.~H., Tyler, G.~L., 1994. Radio occultation studies of the venus
  atmosphere with the magellan spacecraft: 1. experimental description and
  performance. Icarus 110~(1), 71--78.

\bibitem[{STP-TS-012-1(2012)}]{asmestandards2012}
STP-TS-012-1, 2012. Thermophysical {P}roperties of {W}orking {G}ases {U}sed in
  {W}orking {G}as {T}urbine {A}pplications. Standard, ASME Standards
  Technology, LLC, New York.

\bibitem[{Team and Seiff(1987)}]{Seiff1987}
Team, V. B.~S., Seiff, A., 1987. Further information on structure of the
  atmosphere of venus derived from the vega venus balloon and lander mission.
  Advances in space research 7~(12), 323--328.

\bibitem[{Tillner-Roth(1993)}]{tillner1993}
Tillner-Roth, R., 1993. Die thermodynamischen eigenschaften von r152a, r134a
  und ihren gemischen. Forschungsbericht des Deutschen K{\"a}lte-und
  Klimatechnischen Vereins No. 41.

\bibitem[{{Von Zahn} et~al.(1983){Von Zahn}, {Kumar}, {Niemann}, and
  {Prinn}}]{VonZahn1983}
{Von Zahn}, U., {Kumar}, S., {Niemann}, H., {Prinn}, R., 1983. {Composition of
  the Venus Atmosphere}. University of {A}rizona {P}ress, pp. 299--430.

\bibitem[{{Zasova} et~al.(2006){Zasova}, {Moroz}, {Linkin}, {Khatuntsev}, and
  {Maiorov}}]{Zasova2006}
{Zasova}, L.~V., {Moroz}, V.~I., {Linkin}, V.~M., {Khatuntsev}, I.~V.,
  {Maiorov}, B.~S., July 2006. {Structure of the Venusian atmosphere from
  surface up to 100 km}. Cosmic Research 44, 364--383.

\end{thebibliography}

\end{document}